\documentclass[fleqn,12pt]{wlscirep}

\usepackage[utf8]{inputenc}
\usepackage[T1]{fontenc}
\usepackage{url}
\usepackage{graphicx}
\usepackage{booktabs}
\usepackage{subcaption}
\usepackage{hyperref}
\usepackage{amsmath}
\usepackage{amsfonts}
\usepackage{algorithmic,algorithm}

\usepackage{gensymb}
\usepackage{soul}
\usepackage{color}
\usepackage{longtable}
\usepackage{multirow}
\newcommand{\reasat}{\textcolor{red}}

\usepackage{amssymb}

\title{\textbf{MSTT-199}: MRI Dataset for \underline{M}usculoskeletal \underline{S}oft \underline{T}issue \underline{T}umor Segmentation}

\author[a]{Tahsin Reasat}
\author[b]{Stephen Chenard}
\author[b]{Akhil Rekulapelli}
\author[c]{Nicholas Chadwick}
\author[c]{Joanna Shechtel}
\author[c]{Katherine van Schaik}
\author[a,c,d]{David S.~Smith}
\author[b]{Joshua Lawrenz}

\affil[a]{Department of Electrical and Computer Engineering, Vanderbilt University, Nashville, 37235, Tennessee, USA}
            
\affil[b]{Department of Orthop{\ae}dic Surgery,Vanderbilt University Medical Center, Nashville, 37232, Tennessee, USA}

\affil[c]{Department of Radiology and Radiological Sciences,Vanderbilt University Medical Center, Nashville, 37232, Tennessee, USA}

\affil[d]{Institute of Imaging Science, Vanderbilt University Medical Center, Nashville, 37232, Tennessee, USA}

\begin{abstract}

Accurate musculoskeletal soft tissue tumor segmentation is vital for assessing tumor size, location, diagnosis, and response to treatment, thereby influencing patient outcomes. However, segmentation of these tumors requires clinical expertise, and an automated segmentation model would save valuable time for both clinician and patient. Training an automatic model requires a large dataset of annotated images. In this work, we describe the collection of an MR imaging dataset of 199 musculoskeletal soft tissue tumors from 199 patients. We trained segmentation models on this dataset and then benchmarked them on a publicly available dataset. Our model achieved the state-of-the-art dice score of 0.79 out of the box without any fine tuning, which shows the diversity and utility of our curated dataset. We analyzed the model predictions and found that its performance suffered on fibrous and vascular tumors due to their diverse anatomical location, size, and intensity heterogeneity. The code and models are available in the following github repository \href{https://github.com/Reasat/mstt}{[github]}.

\end{abstract}

\begin{document}
\flushbottom
\maketitle

\thispagestyle{empty}

\section{Introduction}

A musculoskeletal soft tissue tumor (MSTT) is an abnormal growth or mass that develops within the 
soft tissues of the body that support and connect the musculoskeletal system \cite{goldblum2013enzinger}. Soft tissues encompass a 
variety of structures, including muscles, tendons, ligaments, fat, blood 
vessels, nerves, and connective tissues. MSTTs can arise from any 
of these tissues and can be either benign (non-cancerous) or malignant 
(cancerous).
Benign MSTTs, such as lipomas or fibromas, typically grow slowly 
and do not invade surrounding tissues or metastasize to other parts of the body. 
Malignant MSTTs, on the other hand, can be aggressive and have the 
potential to spread to nearby organs or distant sites, posing a more significant 
health risk.
\cite{goldblum2013enzinger}

Tumor segmentation allows for precise delineation of the tumor boundaries, 
providing accurate measurements of size and shape. This information is 
essential for disease staging and determining the appropriate treatment 
strategy \cite{cormier2004soft}. Moreover, precise segmentation facilitates the 
monitoring of tumor progression and response to therapy over time, enabling 
clinicians to make timely adjustments to the treatment plan 
\cite{cormier2004soft}. Additionally, segmentation enables the identification of 
heterogeneous regions within the tumor allowing diagnosis of varying levels of 
malignancy \cite{cormier2004soft}. 

A crucial step of building an automated model 
that identifies benign and malignant tumors is the manual segmentation of the 
tumor \cite{lawrenz2022can,cui2022multi,fields2021whole,lee2023ensemble}. 
Segmentation is challenging, as tumor appearance can vary in shape, intensity, 
and tissue composition \cite{cormier2004soft}. Additionally, the presence of 
artifacts such as noise, motion, and magnetic susceptibility can further obscure 
tumor boundaries, making segmentation challenging.  Moreover, the lack of 
standardized protocols for acquiring MRI data, leading to variations in image
contrast and quality across different institutions and scanners adds to the 
difficulty.
Even if the clinician has sufficient expertise, manual delineation of tumors in three dimensions is 
a time-consuming process, taking up to half an hour per 
MRI volume \cite{dionisio2021manual}.

In recent times, researchers have strived to automate the MSTT
segmentation process by employing various classical machine learning 
\cite{blackledge2019supervised} and deep learning  \cite{bi2024hybrid, 
neubauer2020soft, diao2021efnet} based methods. Researchers have explored models 
that take both single and multimodal images (MRI, PET, CT scans) as input and 
predict tumor segmentation. These models have been trained and evaluated using a 
small dataset of 51 patients presented in \cite{vallieres2015radiomics}.
The progress of automatic MSTT 
segmentation models has lagged due to the unavailability of large diversified datasets. 
To address this problem we have curated a dataset and trained a segmentation 
model using the data. The contributions of this paper are fourfold:
\begin{itemize}
	\item We created an MSTT segmentation dataset with 199 patients 
		and plan to make it publicly available for future research;
	\item We described our process of selecting the patients, setting up the 
		labeling platform, the annotation protocol, and the curation method;
	\item We created a segmentation model based on the curated data which 
		achieves state-of-the-art (SOTA) result on the only available public 
		dataset and analyzed the results; and
	\item We identify easy versus hard-to-detect types of tumor and make 
		suggestions for future model development as well as data collection.
		% (\reasat{If possible}). 
\end{itemize}

The paper is organized in the following sections. Section \ref{sec_dataset} 
contains the detailed creation of the dataset. Section \ref{sec_model} includes 
the architectural explanation of the segmentation models used in this work. 
Section \ref{section_exp} contains the experimental setup and result analysis.
And finally the paper is concluded in Section \ref{sec_conclusion}.

\section{MSTT-199: Dataset Description}
\label{sec_dataset}

In this section, we describe the 
process of patient selection,  data annotation, annotation protocol, and data 
curation method.

\subsection{Patient Selection}

After receiving institutional review board approval, we queried our 
institution's orthopaedic oncology registry, which includes all patients treated 
for an MSTT at our institution since 1987. Using 
this registry, we initially identified 2,639 patients who underwent definitive 
oncological resection at Vanderbilt University Medical Center and had one of the 
following diagnoses on final pathological review: schwannoma (benign nerve 
tumor), MPNST (malignant peripheral nerve sheath tumor), well-differentiated liposarcoma (benign fat tumor), dedifferentiated liposarcoma (malignant fat tumor), desmoid fibromatosis (benign fibrous tumor), undifferentiated pleomorphic sarcoma 
(malignant fibrous tumor), hemangioma or arteriovenous malformation (benign 
vascular tumor), angiosarcoma (malignant vascular tumor), myxoma (benign myxoid 
tumor), or myxoid fibrosarcoma (malignant myxoid tumor). The tissue type of the tumors were divided in five broad categories Fibrous, Fat, Myxoid, Nerve, and Vascular.

We retrospectively 
reviewed the electronic health records of patients in reverse chronological 
order and sequentially included patients who had a tumor with largest dimension 
greater than 3 cm, and a pre-operative MRI that included both an axial T1 and an 
axial T2 fat-saturation sequence. Patients with MRI sequences that were deemed 
to be incomplete or of poor image quality as determined by a board-certified 
orthopedic oncologist or musculoskeletal radiologist were excluded. Records were 
reviewed until $\sim$40 eligible patients with each tumor tissue type (Fibrous, Fat, 
Myxoid, Nerve, Vascular) were included, or all available records were exhausted. 
This resulted in a collection of 199 patients with 199 MSTTs. 
% Forty of these patients were collected from an earlier study on myxoids 
% \cite{lawrenz2022can}. 

% \reasat{Demographic and clinical data were abstracted from the electronic health record and are presented in table ***}
% \reasat{Need help regarding patient selection from Stephen, Akhil}
% \reasat{Report distribution of age, gender, acquisition period, scanner info, cohort source, scanning resolution, various imaging information}
% \reasat{Report metastases information}
% \reasat{Patient selected based on some criteria, tried to do manual registration, if not matched Stephen+Akhil looked for more}
% \reasat{make distribution of malignant, benign, take help from Stephen, Akhil}

\subsection{Labeling Platform}

We selected LabelStudio \cite{LabelStudio} as our data annotation platform. 
LabelStudio provides a web browser-based labeling platform that has minimal user 
setup overhead at the annotator's end. Radiologists can log in through the 
provided URL, create an account, and start annotating the assigned image. 
LabelStudio provides a customizable user management system through which user 
activity and dataset progress can be tracked with ease. 

\begin{figure}
	\centering
	\includegraphics[ 
	width=0.5\columnwidth]{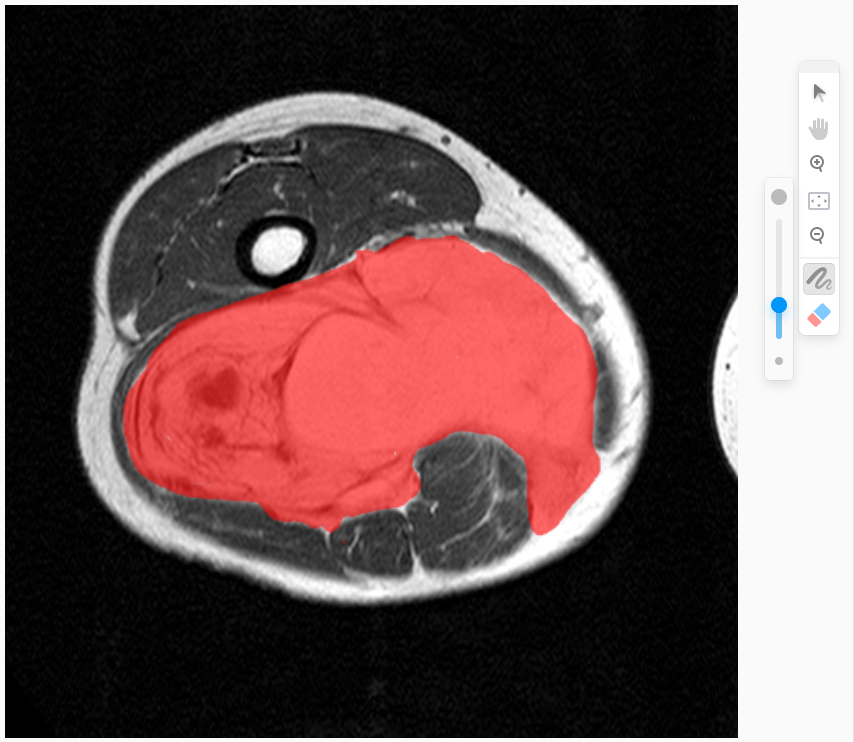}
	\caption{The Label Studio annotation setup used for this project. The annotators can choose a brush size 
	and annotate the tumor at different granularity.}
	\label{fig:labelstudio}
\end{figure}

Both T1- and T2-weighted images were present for each tumor and registered 
using ANTs \cite{avants2011reproducible} algorithm on the 3D Slicer software 
\cite{fedorov20123d}. The smaller image of the pair was also resampled to match 
the resolution of the largest image. Following registration, the axial slices of 
MRI images were exported to PNG files and listed on the project along with 
necessary metadata such as patient ID, modality, type of tumor, slice instance 
number, anatomy, etc. While logged in to the platform, the annotator could 
search images via patient ID and modality and sort by instance number. After 
selecting an image, the annotator drew segmentation masks on top of the tumor 
area using a brush tool. A screenshot of the annotation user interface is shown 
in Fig.~\ref{fig:labelstudio}. The size of the brush tool can be varied for fine 
or coarse masks. The choice of brush size creates a trade-off between accuracy 
and annotation time. 
% \reasat{for each step add a screenshot from LabelStudio}.

\subsection{Annotation Protocol}

% \reasat{How did we define the tumor area to annotate, who annotated the image, 
% how did we solve confusions that needed to be addressed}
% We annotated 163 MRI images (from 163 patients) with soft tissue tumors. 
Annotations were done primarily on the T2-weighted images except for the Fat tumors which were mostly done on the T1-weighted images due to better tumor visibility. The axial slices of the images were uploaded on a privately deployed annotation platform called LabelStudio \cite{tkachenko2020label}. The annotators logged into the platform to annotate and submit their annotations. 

The data was annotated in three stages. In the first stage, a radiologist (K) 
annotated the center slice of the tumor present in each MRI. In the second 
stage, three annotators (S, A, T
%,\reasat{maybe some words on qualifications?} 
)
annotated the adjacent slices of the annotation following the guiding 
annotations done by K. The questions or confusion that arose during the second 
stage were all mitigated via discussion with the radiologist.

In the final stage, the complete annotations were sent to the radiologists, J, 
K, and N 
% \reasat{(maybe some words on expertise?)}
for final review. The patients were randomly divided among the three 
radiologists and had one radiologist review per image. 

The radiologists identified tumor area regions by looking at T2 signal hyperintensity or edema on fluid-sensitive sequences and also using the contrast between normal muscle or tumors and surrounding fat planes on the conventional T1 sequences. While annotating most of the central mass is demarcated; however, some tumors may have narrow tail-like extension from the margins of the mass which might be excluded. Radiologists took image noise and artifact into consideration during the interpretation of the MR images and discounted them without difficulty.

\subsection{Dataset Statistics}
\begin{figure}
    \centering
    \begin{subfigure}[t]{0.3\textwidth}
    \includegraphics[width=\columnwidth]{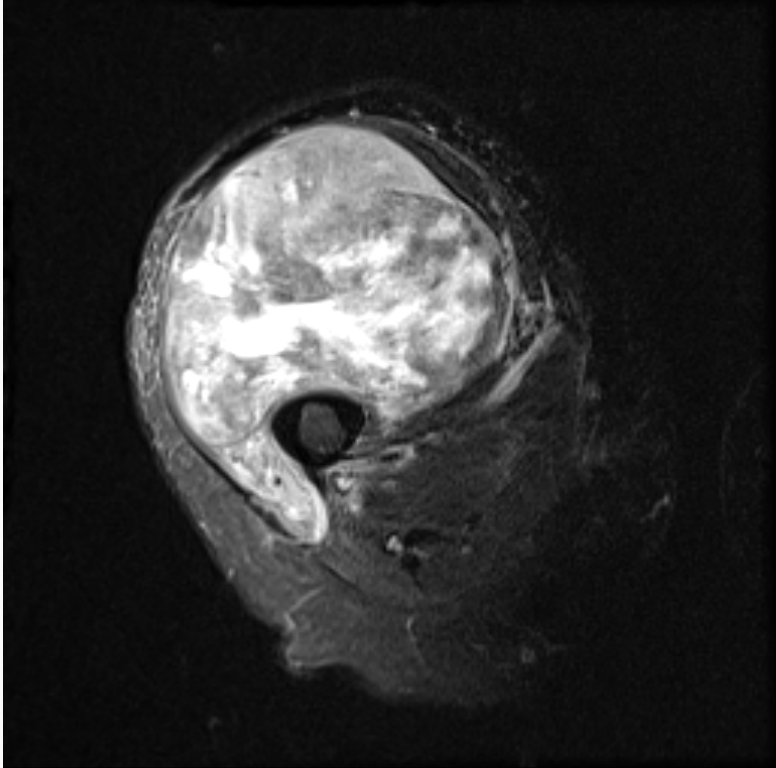}
    \caption*{Fibrous}
    \end{subfigure}
    \begin{subfigure}[t]{0.3\textwidth}
    \includegraphics[width=\columnwidth]{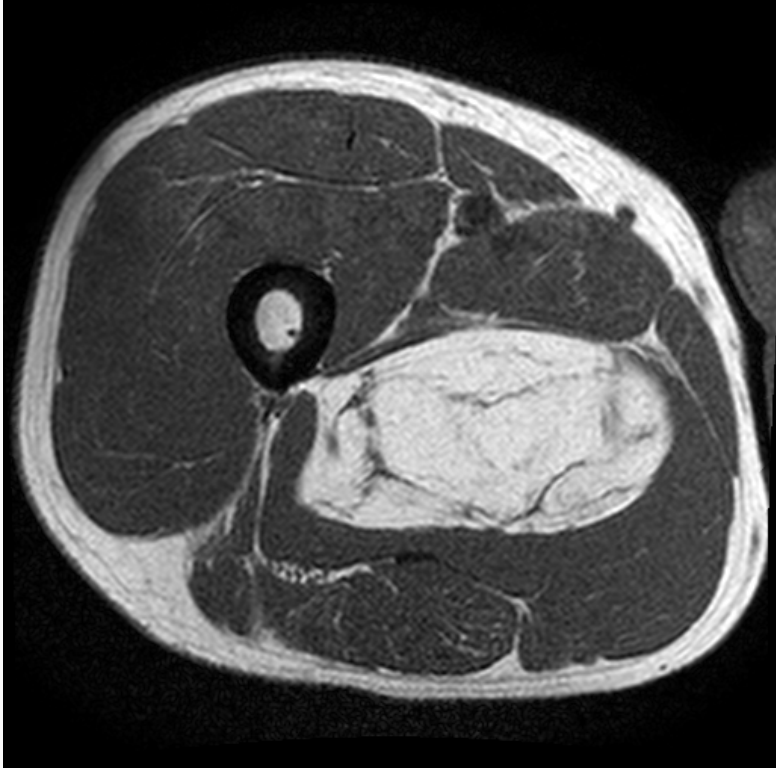}
    \caption*{Fat}
    \end{subfigure}
    \\
    \begin{subfigure}[t]{0.285\textwidth}
    \includegraphics[width=\columnwidth]{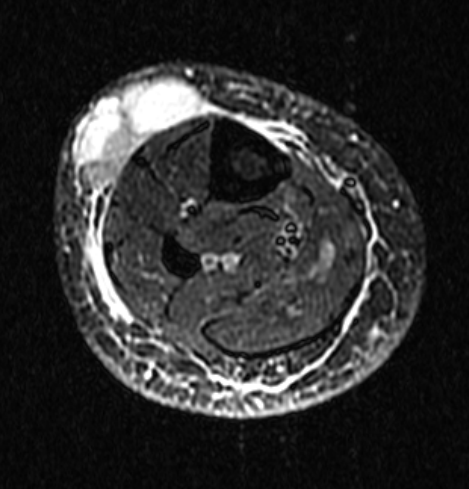}
    \caption*{Myxoid}
    \end{subfigure}
    \begin{subfigure}[t]{0.3\textwidth}
    \includegraphics[width=\columnwidth]{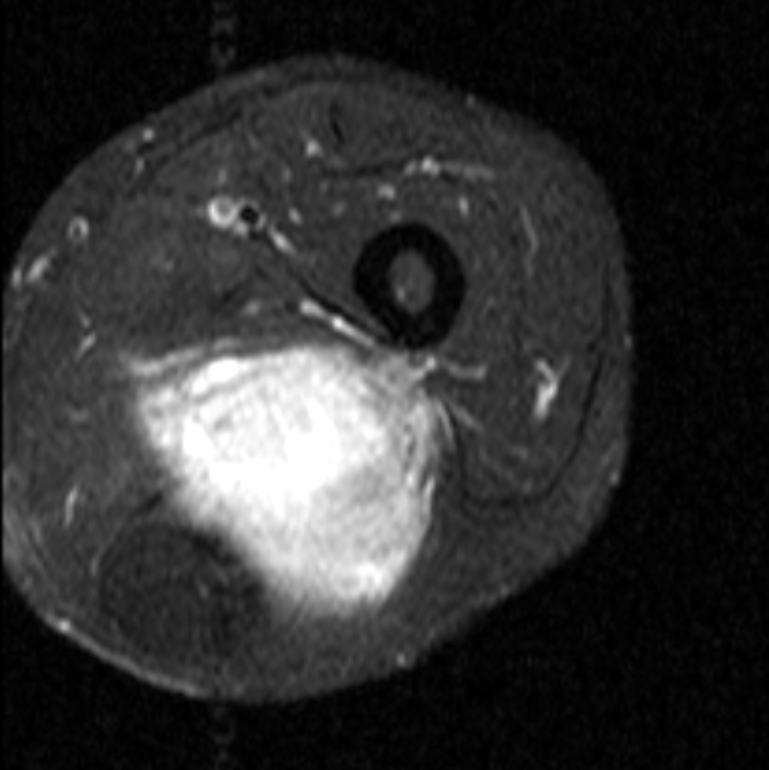}
    \caption*{Nerve}
    \end{subfigure}
    \begin{subfigure}[t]{0.3\textwidth}
    \includegraphics[width=\columnwidth]{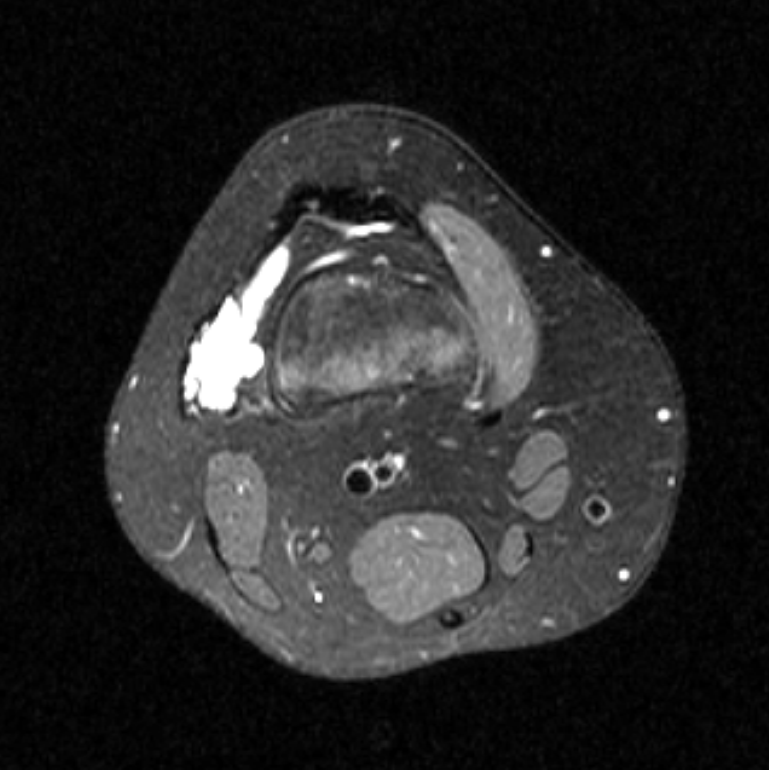}
    \caption*{Vascular}
    \end{subfigure}
    \caption{Example of different tissue types present in the MSTT-199 corpus. The Fat tumor is brighter on T1 while the rest of the tissue types are brighter on T2.}
    \label{fig:tissue_examples}
\end{figure}

% \reasat{figure, showing class distribution, anatomy distribution,  tumor size distribution, image size distribution, scan resolution distribution, image intensity distribution, }
\begin{table}[]
\centering
\begin{tabular}{lcccccc}
\toprule
                               & \multicolumn{5}{c}{Tissue Type}                                                                                       &                           \\
                               \midrule
Tumor Site & Fibrous & Fat & Myxoid & Nerve & Vascular & Total \\
\midrule
thigh                          & 21                    & 35                    & 26                    & 10                    & 11                    & 103                       \\
leg                            & 5                     & 2                     & 2                     & 11                    & 5                     & 25                        \\
glute                          & 0                     & 2                     & 11                    & 3                     & 3                     & 19                        \\
forearm                        & 1                     & 0                     & 0                     & 4                     & 8                     & 13                        \\
arm                            & 1                     & 1                     & 0                     & 8                     & 0                     & 10                        \\
shoulder                       & 3                     & 0                     & 0                     & 0                     & 4                     & 7                         \\
pelvis                         & 1                     & 0                     & 0                     & 1                     & 2                     & 4                         \\
hand                           & 0                     & 0                     & 0                     & 1                     & 3                     & 4                         \\
foot                           & 1                     & 0                     & 0                     & 1                     & 1                     & 3                         \\
chest wall                     & 2                     & 0                     & 1                     & 0                     & 0                     & 3                         \\
neck                           & 2                     & 0                     & 0                     & 0                     & 1                     & 3                         \\
abdominal wall                 & 2                     & 0                     & 0                     & 0                     & 0                     & 2                         \\
axilla                         & 0                     & 0                     & 0                     & 1                     & 0                     & 1                         \\
back                           & 1                     & 0                     & 0                     & 0                     & 0                     & 1                         \\
flank                          & 0                     & 0                     & 0                     & 0                     & 1                     & 1\\
\midrule
Total & 40 & 40 & 40 & 40 & 39 & 199\\
\midrule
\end{tabular}
\caption{Tissue type and anatomical distribution of the tumors. }
\label{table:anatomy_tissue_distribution}

\end{table}

\begin{figure}
    \centering
    \includegraphics[width=0.7\columnwidth]{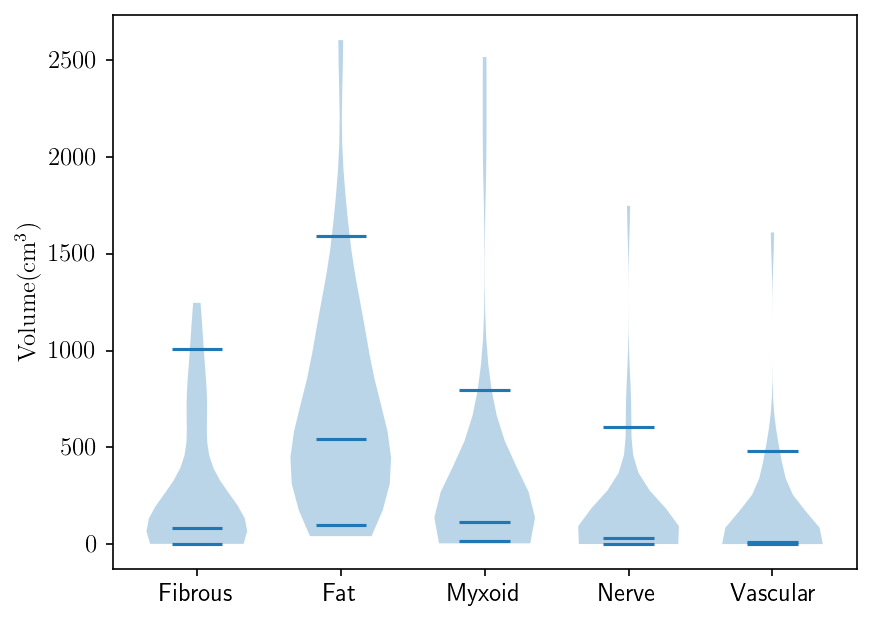}
    \caption{Tumor volume distribution. Fat tumors were on average larger than those for the other subtypes, while the other four subtypes were more similar.
    % \reasat{Should I plot with outliers}
    }
    \label{fig:tumor_size}
\end{figure}

\begin{figure}
    \centering
    \includegraphics[width=0.7\columnwidth]{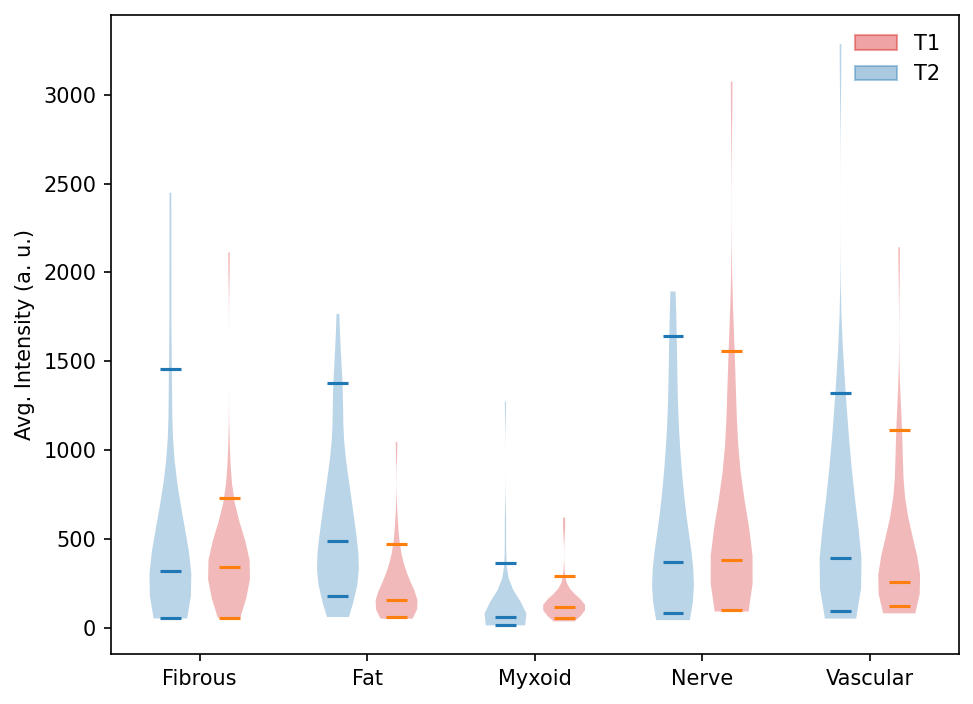}
    \caption{Average tumor intensity distribution in T1 and T2 image. Myxoid tumors tended to have a lower intensity on average on both T1 and T2, while the other tumors were more similar in intensity. 
    % \reasat{Should I plot with outliers?}
    }
    \label{fig:tumor_intensity_dist}
\end{figure}

The tissue type statistics and anatomy distribution in the dataset are shown in 
Table \ref{table:anatomy_tissue_distribution}, and examples of tissue images are 
shown in Fig.~\ref{fig:tissue_examples}. Although we assembled a balanced 
dataset for each tissue type, it was difficult to keep the anatomical location 
distribution balanced. MSTTs are most prevalent in the extremities 
with very few occurrences in the trunk or head and neck region 
\cite{cormier2004soft, morrison2003soft}.
% Although STT affects connective tissues throughout the body, the most common 
% location is in the extremities, in 59\% of cases, followed by the trunk, in 
% 19\%, the retroperitoneum, in 15\%, and the head/neck region, in 9\% 
Fig.~\ref{fig:tumor_size} shows the tumor size variation for different 
tissue types. The Fat tumors are usually larger in size, and the nerve and 
vascular tumors are comparatively smaller.
Fig.~\ref{fig:tumor_intensity_dist} shows the intensity distribution across the 
tissue types. For fibrous, nerve, and vascular tissue types there is an overlap 
in the distribution of the intensities from the two modalities. However, Fat 
tumors show higher intensity on T1 (Fig.~\ref{fig:tissue_examples}),  whereas, the myxoid tumors show a higher intensity 
distribution on T2. This justifies our choice to annotate the fat 
tumors on T1 modality and the myxoid tumors on T2 modality.

% \begin{figure}
%     \centering
%     \includegraphics[width=0.7\columnwidth]{fig/tumor_size_dist.png}
%     \caption{Tumor size distribution. \reasat{Should I plot without outliers}}
%     \label{fig:tumor_size}
% \end{figure}

% \begin{figure}
%     \centering
%     \includegraphics[width=0.7\columnwidth]{fig/tumor_intensity_dist.png}
%     \caption{Tumor intensity distribution. \reasat{should I plot without outliers?}}
%     \label{fig:tumor_intensity_T1}
% \end{figure}

\section{Segmentation Models}
\label{sec_model}
In this section, we describe the segmentation model architectures and the loss function used to train the model.

\subsection{Architecture}

\begin{figure}
    \centering
    \begin{subfigure}[t]{0.7\textwidth}
    \includegraphics[width=\columnwidth]{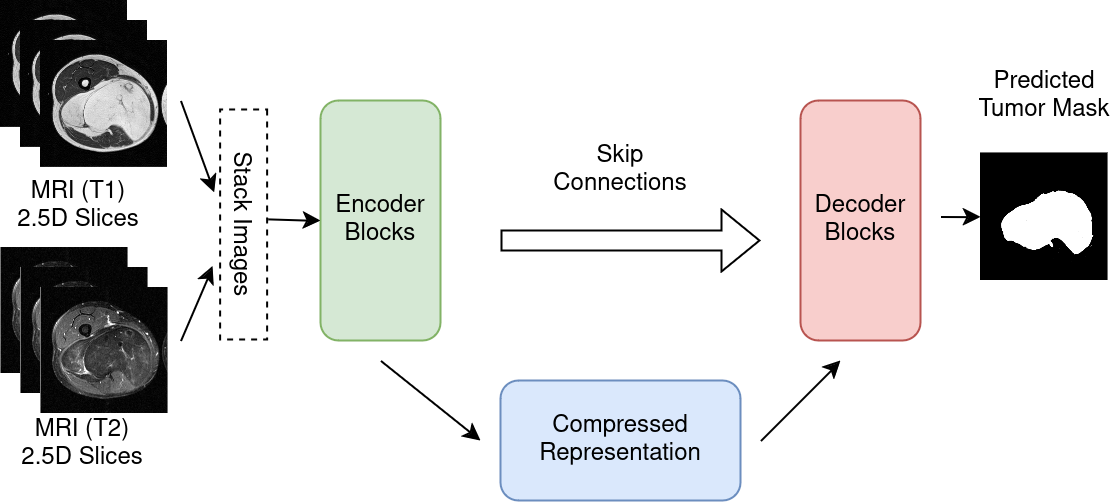}
    \caption{U-Net}
    \label{fig:unet}
    \end{subfigure}\\
    \vspace{1mm}
    \begin{subfigure}[t]{0.7\textwidth}
    \includegraphics[width=\columnwidth]{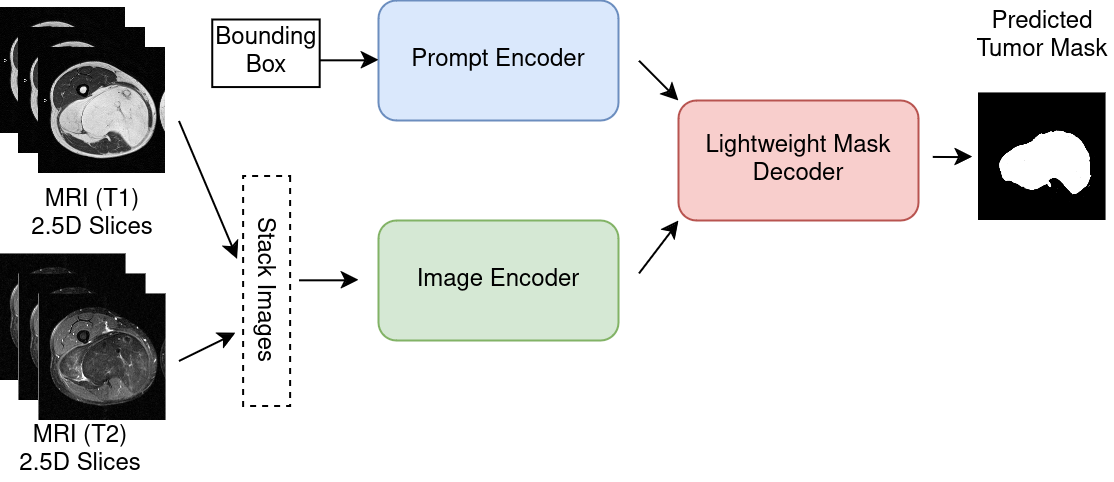}
    \caption{Segment Anything Model (SAM)}
    \label{fig:sam}
    \end{subfigure}

	\caption{The segmentation model architectures. The input to both the 
	segmentation models is a multi-modal MRI image (T1 and T2). The 2.5-D slices 
	coming from each modality are stacked to create a six-channel input. 
	\ref{fig:unet}) The U-Net segmentation model has a U-shaped structure due to 
	skip connections going from the encoders to the decoders. \ref{fig:sam}) The 
	SAM model has an additional prompt encoder. For automatic segmentation, the 
	prompt encoder receives a bounding box drawn over the full image.}
   
\end{figure}

% \reasat{We used a 2.5D approach in training the model. Three consecutive slices of MRI images were selected as a training sample and the model was tasked to predict tumor content on the center sample.}

% \reasat{encoder chosen, augmentations, dice score definition}
We used the U-Net \cite{ronneberger2015u} architecture as our segmentation model.
The U-Net segmentation model is a convolutional neural network architecture 
initially designed for biomedical image segmentation tasks 
\cite{ronneberger2015u}. U-Net has two main components: an encoder and a decoder. 
The encoder downsamples the input image in steps and reduces it to a 
representation rich in information but with the loss of spatial resolution. The 
decoder takes this representation and gradually upsamples it to reconstruct a 
probability output having the spatial size of the image. The output map is 
thresholded to create a binary mask of the foreground object. There are shortcut 
connections (or skip connections) between different stages of the encoder and 
decoder that have similar feature dimensions which enables information flow 
between feature blocks. The encoder, decoder, and shortcut connections create a 
U-shaped visualization for the model (Fig.~\ref{fig:unet}), hence the name U-Net.  

\subsubsection{Encoder}
The encoder consists of a series of convolutional blocks. Each block typically 
consists of two consecutive convolutional layers followed by a rectified linear 
unit (ReLU) \cite{fukushima1969visual} activation function and a max-pooling 
layer. The purpose of this path is to capture context and spatial information 
from the input image while reducing its spatial dimensions. As the network 
progresses through the contracting path, the receptive field increases while the 
spatial resolution decreases.

\subsubsection{Decoder}
The decoder is responsible for upsampling the feature maps to the original input 
resolution. Each block in this path consists of an upsampling operation (usually 
transposed convolution or interpolation), followed by concatenation with feature 
maps from the encoder path, and then a series of convolutional layers. The 
concatenation operation helps in preserving fine-grained details from the 
encoder path, facilitating precise segmentation.

\subsubsection{Shortcut Connections}
One of the key features of U-Net is the skip connections that directly connect 
corresponding layers between the encoder and the decoder. These connections 
allow the network to bypass the loss of spatial information during downsampling 
and aid in the precise localization of objects in the segmentation masks. Skip 
connections provide a shortcut for gradient flow during training, helping to 
mitigate the vanishing gradient problem and enabling faster convergence.

\subsubsection{Final Layer}
The final layer of the network consists of a $1\times 1$ convolutional layer 
followed by a sigmoid activation function. This layer produces the segmentation 
mask with the same spatial dimensions as the input image, where each pixel 
represents the predicted class or label. 

\subsection{Segment Anything Model (SAM)}
The segment anything model (SAM) \cite{kirillov2023segment} is a foundational 
model proposed for image segmentation. It has been trained on a large dataset of 
diverse images and can be applied to a wider range of segmentation tasks. This 
model has been further fine tuned on a large-scale medical image segmentation 
dataset with 1,570,263 medical image-mask pairs, covering 10 imaging modalities, 
over 30 cancer types, and a multitude of imaging protocols to create the MedSAM 
model \cite{ma2024segment}.  The SAM model has three components, illustrated in 
Fig.~\ref{fig:sam}: an image encoder, a flexible prompt encoder, and a fast mask 
decoder. These components are described at a high level here.

\subsubsection{Image Encoder}
SAM utilizes a pre-trained Vision Transformer (ViT) \cite{dosovitskiy2020image} 
that is adapted to process high-resolution inputs. This encoder runs once per 
image and can efficiently process the image before any prompting.

\subsubsection{Prompt Encoder} In the original SAM model, two types of 
prompts are considered: sparse (points, boxes, text) and dense (masks). These 
prompts are transformed into an embedding and combined element-wise with the 
image embedding. To make the model fully automated it can be supplied with the 
bounding box of the full image and the model predicts masks for all available 
foreground objects.

\subsubsection{Mask Decoder}
The mask decoder takes the image embedding and the prompt embeddings to produce 
a mask. It employs a modified transformer decoder block which uses prompt 
self-attention and cross-attention to update all embeddings 
\cite{vaswani2017attention}. The output of these attention layers is upsampled, 
and a fully connected layer maps the output token to a dynamic linear 
classifier, which computes the mask foreground probability at each image 
location. 
% \reasat{check details}

\subsection{Loss}
Segmentation models are typically trained using stochastic gradient descent which 
optimizes a loss function computed over the final layer.
We used binary cross entropy as our loss function:
% \reasat{bce+dice did not improve score}
\begin{align}
	%\mathcal{L}_{dice} = & 1 - \frac{2TP}{2TPFP+FN+eps}\\
	\mathcal{L}_\mathrm{bce} = & y \ln(p) + (1-y) \ln(1-p),\ y\in\{0, 1\}.
	%\mathcal{L} = & \mathcal{L}_{dice} + \mathcal{L}_{bce} 
\end{align}

\section{Experiment and Result Analysis}
\label{section_exp}
In this section, we explain the details of the used datasets, along with the 
parameters for data preprocessing and model training. Additionally, we analyze 
the results produced from the experiments.

\subsection{External Dataset}
Alongside evaluating the trained models on the test partition of our dataset, we 
performed out-of-domain evaluation on a publicly available soft tissue sarcoma 
dataset (STS)\cite{vallieres2015radiomics}, accessible via The Cancer Imaging 
Archive \cite{clark2013cancer}. This dataset consisted of 51 patients with 
sarcomas located in the extremities, sourced from various sites and scanners, 
resulting in high heterogeneity. Each patient's data contained four different 
imaging modalities, including two paired MRI scans (T1 and T2) and a PET/CT 
scan. The MRI and PET/CT scans were conducted on different days, leading to 
variations in body positioning and anatomy. Tumor annotations are already 
provided in the dataset, delineated on the T2 scans. Additionally, we 
co-registered each T1 image onto the corresponding T2 image using the ANTs 
\cite{avants2011reproducible} algorithm in 3D Slicer. However, while processing 
the dataset we found two image pairs 
%(STS\_025, STS\_029)
with excessive movement between the modalities that were excluded in our 
analysis. The processed STS data in the 3D NIfTI format is publicly available at \url{provide url}.

\subsection{Experimental Details}

The 3D MRI images were resampled to have a voxel size of 1 mm $\times$ 1 mm 
$\times$ 1 mm. The intensity values of each image were clamped between 0.05\% and 
99.95\% of the distribution of that particular image intensities to exclude 
outliers. The images were normalized using the min-max normalization procedure 
where the minimum and maximum intensities were image dependent.  In the axial, 
sagittal, and coronal direction of the volume, 3 consecutive slices were grouped 
to create a 2.5D slice. Slices with more than 100 tumor voxels were used to 
train the model. We used 5-fold cross-validation over the 199 patients. 
% \reasat{add information of total slices and slices with tumor after 
% thresholding, also one patient got discarded due to thresholding}.
The input size to the model was 256 $\times$ 256. Padding or cropping was done if 
needed. The augmentations used were random crop, horizontal, vertical flip, 
gamma, brightness, contrast, Gaussian blur, motion blur, and grid distortion. 
% \reasat{adding lr, optimizer info, batch size, epochs, pre-trained weights, 
% also inference details, crop size, tta}
% \reasat{used more complicated MRI-specific augmentations from MONAI, didn't 
% improve score}.

For the U-Net encoder, we used the pre-trained se\_resnext50\_32x4d 
\cite{xie2017aggregated} model and for the SAM image encoder we used the 
LiteMedSam model (a smaller version of the MedSAM model with a similar 
performance) weights which was provide in \cite{ma2024segment}. While training the 
segmentation models we fine tuned both the encoder and the decoder. The 
performance of the models was evaluated using Dice coefficient (Dice), 
i.e.,
\begin{align}
	\mathrm{Dice} = \frac{2\mathrm{TP}}{2\mathrm{TP}+\mathrm{FP}+\mathrm{FN}}
\end{align}.
Here, TP is true positive, FP is false positive, FN is false negative.  
The full details of the training parameters are listed in Appendix \ref{app:training_params}. During inference the images were resized to 256 $\times$ 256. We did test time augmentations during the inference (horizontal and vertical flip, rotated by 90\degree) of each slice and the average output was taken. For each image volume, we conducted inference in the axial, coronal, and sagittal  direction and averaged the output.

\subsection{Result Analysis}

\begin{figure}
    \centering
    \begin{subfigure}[t]{0.40\textwidth}
    \includegraphics[width=\textwidth, trim={0cm 9cm 26cm 17cm},clip]{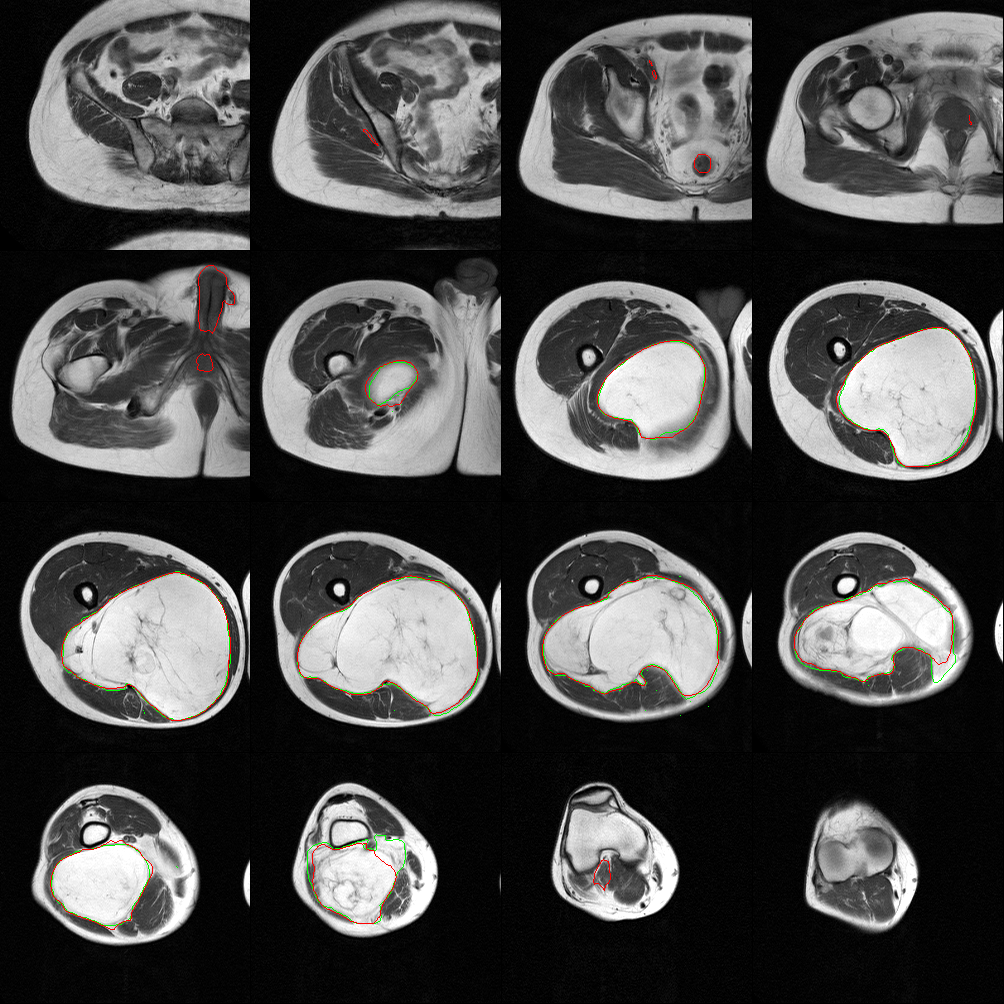}
    %left bottom right top
    \caption*{Large homogeneous Fat tumor in the thigh.}
    \end{subfigure}
    \begin{subfigure}[t]{0.40\textwidth}
    \includegraphics[width=\textwidth, trim={10cm 10cm 18cm 18cm},clip]{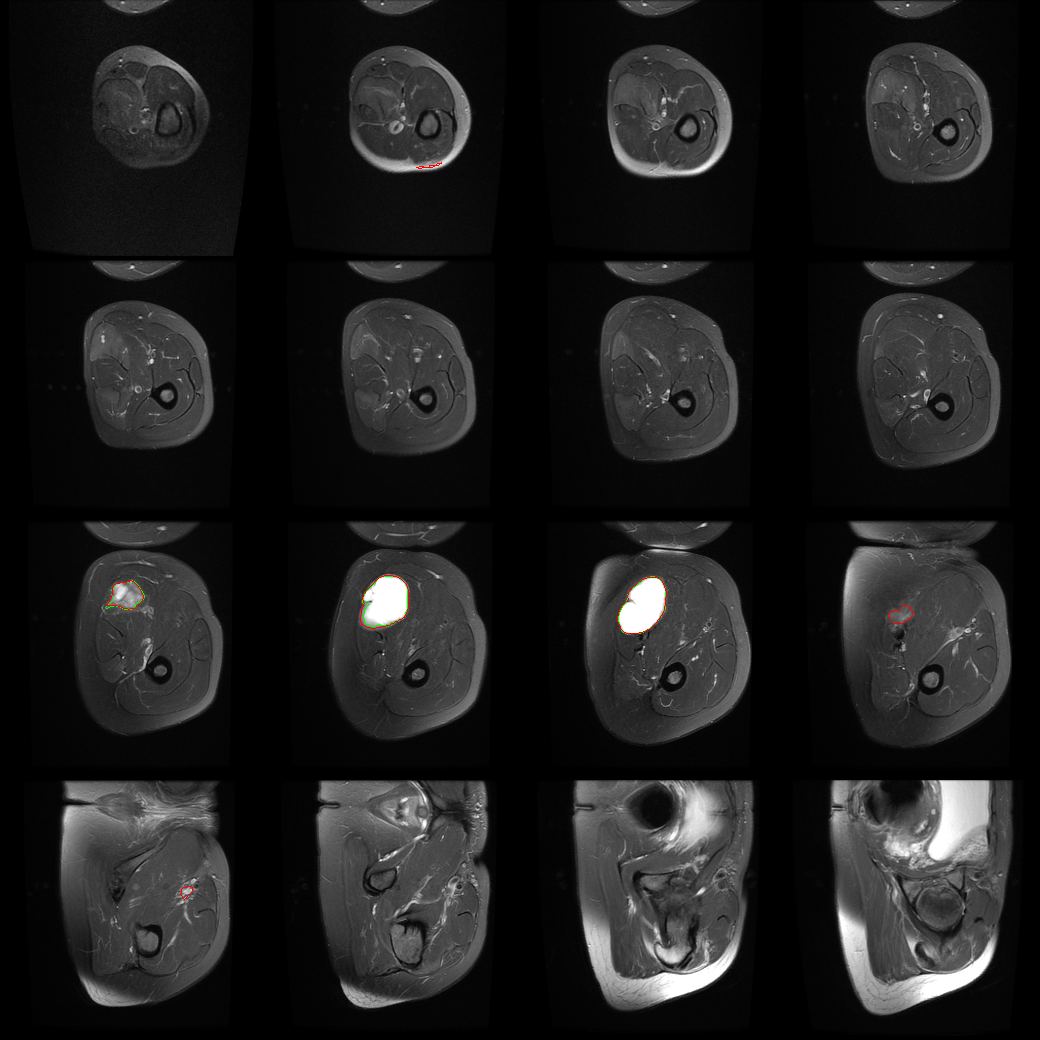}
    %left bottom right top
    \caption*{Medium sized homogeneous myxoid tumor in the thigh.}
    \end{subfigure}\\
    
    \begin{subfigure}[t]{0.40\textwidth}
    \includegraphics[width=\textwidth, trim={16.5cm 9cm 9cm 17cm},clip]{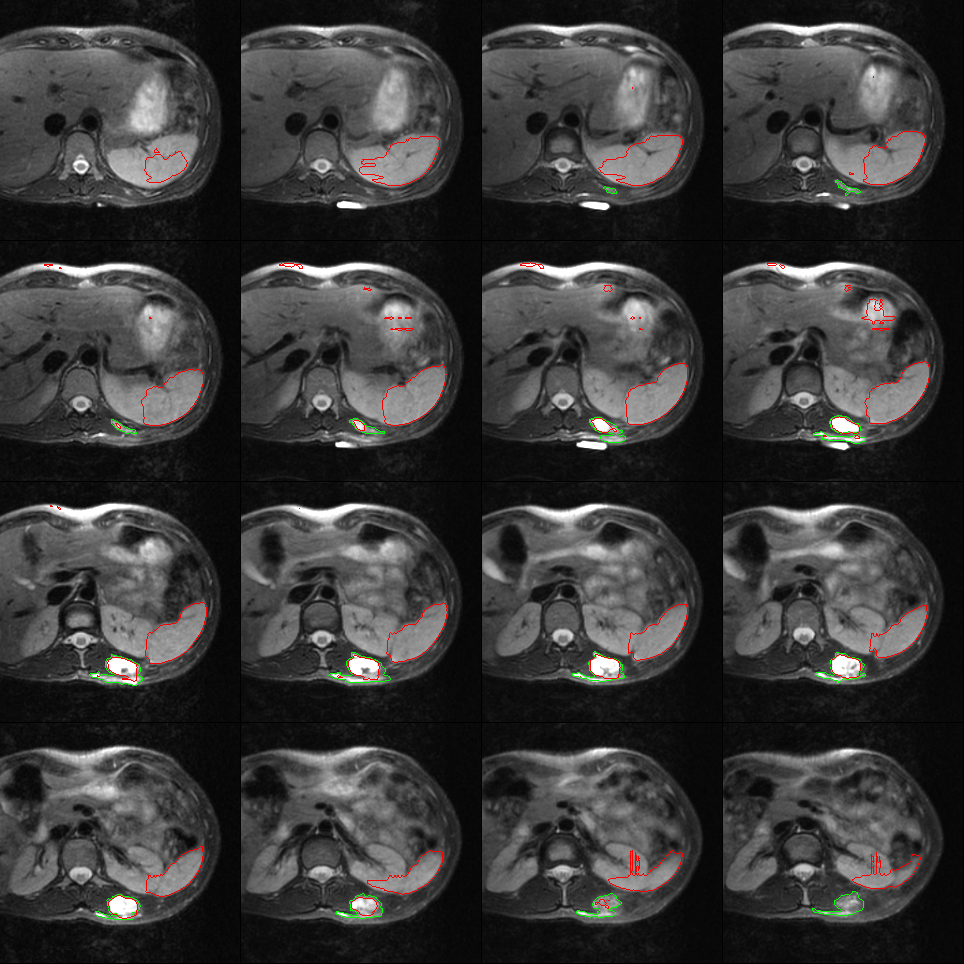}
    %left bottom right top

	\caption*{Small vascular tumor in the presence of confounding tissue 
		structures in the flank (a less represented anatomy).}

    \end{subfigure}
    \begin{subfigure}[t]{0.40\textwidth}
    \includegraphics[width=\textwidth, trim={22cm 11cm 11cm 22.5cm},clip]{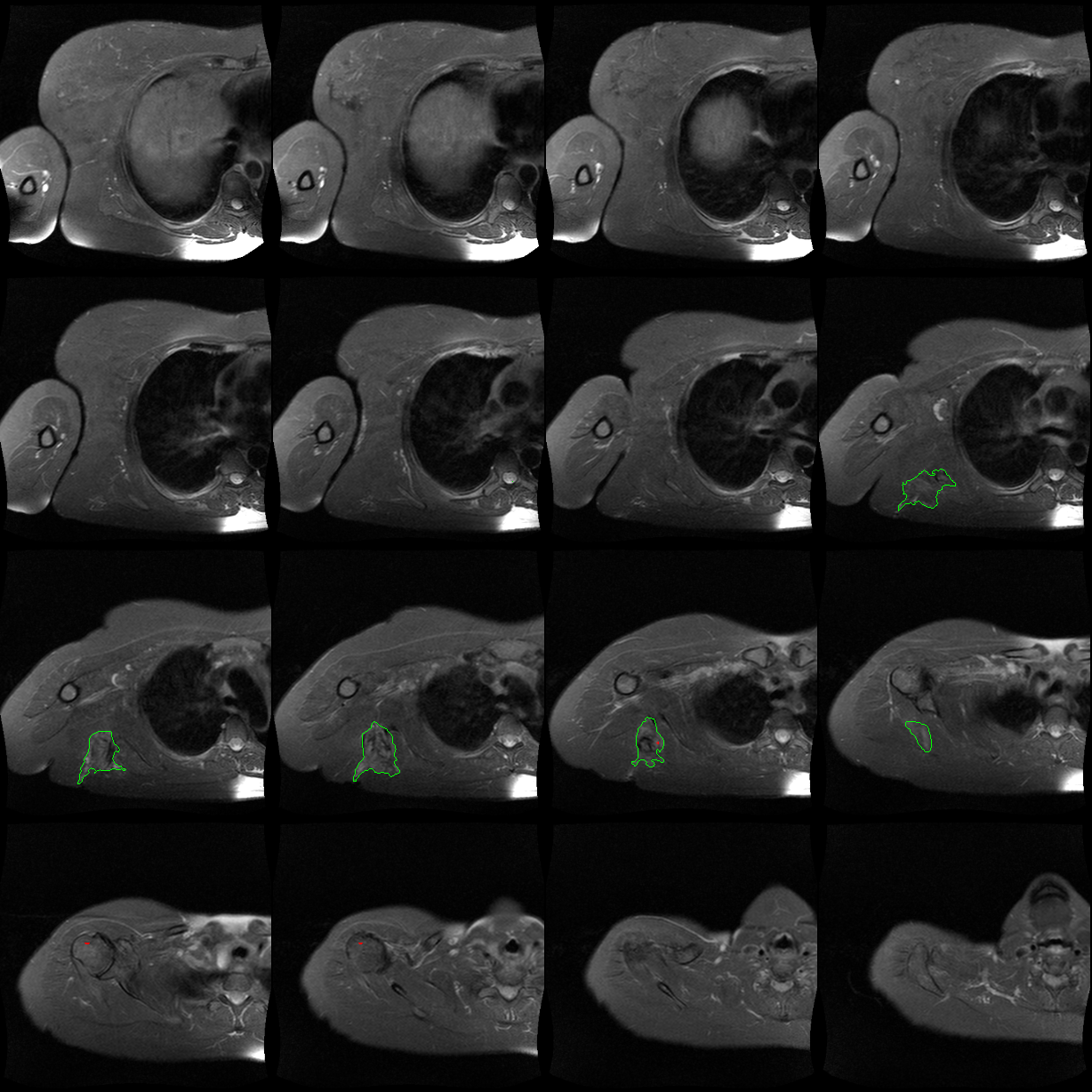}
    %left bottom right top

	\caption*{Small fibrous tissue structure not easily distinguishable from 
		surrounding tissue in the chest wall (another less represented 
		anatomy).}

    \end{subfigure}

	\caption{Examples of model predictions. The top row shows examples where the 
	model successfully draws the contour around the image and the bottom rows 
	show failures. The green contour denotes ground truth and the red contour 
	denotes model predictions.}

    \label{fig:predicted_mask_example}
\end{figure}

% \usepackage{multirow}
\begin{table}[]
\centering
\begin{tabular}{lcccc}
\toprule
\multirow{2}{*}{Model} & \multirow{2}{*}{Input Modality} & \multirow{2}{*}{Train Domain} & \multicolumn{2}{c}{Test Domain} \\
\cline{4-5}
                       &                                 &                               & STS        & MSTT-199       \\
\midrule

U-Net                   & MRI                             & MSTT-199                         & 0.79           & 0.68            \\
LiteMedSAM                    & MRI                             & MSTT-199                         &   0.80         &   0.67          \\
Multi-Branch U-Net\cite{neubauer2020soft}           & MRI, PET                        & STS                           & 0.77           &   -       \\ 
\bottomrule
\end{tabular}

\caption{Performance comparison of the segmentation models trained on the 
	various datasets. The models trained on the MSTT-199 dataset outperform 
	existing benchmarks.}

\label{tab:benchmark}
\end{table}

\begin{table}[htbp]
\centering
\begin{tabular}{lc}
\toprule
Tissue Type                & Dice       \\
\midrule
Fibrous                    & 0.507                        \\
Fat                    & 0.744                         \\
Myxoid                    & 0.823                         \\
Nerve                    & 0.748                         \\
Vascular                    & 0.584                         \\
\midrule

Average                 & 0.681 \\
\bottomrule
\end{tabular}
\caption{Average dice score for different tumor tissue types.}
\label{table:dice_across_tissue}
\end{table}
\begin{figure}
    \centering
    \includegraphics[width=0.7\textwidth]{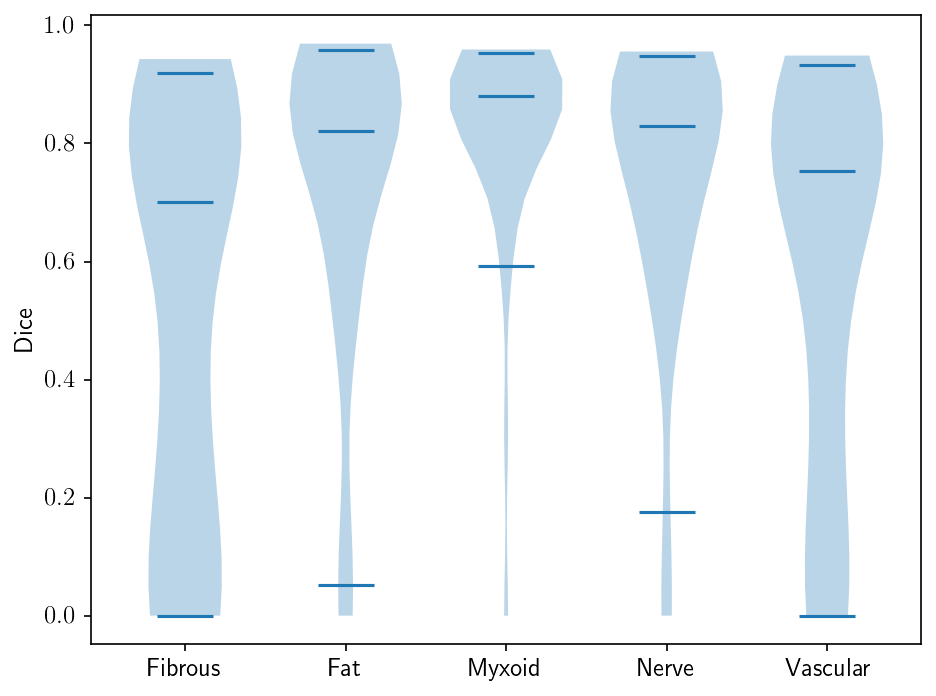}
    \caption{Dice distribution across tissue types.}
    \label{fig:dice_across_tissue}
\end{figure}

In Table \ref{tab:benchmark}, we report the mean dice score achieved in the 
5-fold cross-validation experiment in the MSTT-199 dataset as well as the STS 
dataset and compare it with the existing model in the literature. For the STS 
dataset, there is no predefined test set, so we evaluate our model on the whole 
dataset. The existing Multi-Branch U-Net \cite{neubauer2020soft} model uses a 
5-fold cross-validation approach (which involves domain-specific training). 
Additionally, it uses multiple imaging modalities (MRI and PET) as input. 
Whereas, our simpler U-Net with no domain-specific training, outperforms the 
existing benchmark. This shows the diversity and usefulness of our dataset. 
Lower Dice scores on the MSTT-199 test domain show that our dataset has much harder 
samples compared to STS. Additionally, we observe that the LiteMedSAM model does 
not outperform the U-Net-based model. This is probably because the LiteMedSAM did 
not have an STT segmentation dataset in its large-scale pretraining phase and 
may not have learned features that are related to the STT segmentation task. Additionally, it's pretraining task was a semi-supervised prompting based approach where a tumor bounding box was provided alongside images. Without this additional information the model performance seems to suffer.

In Table \ref{table:dice_across_tissue}, we report the average Dice score 
obtained across tissue types and show the distribution spread using boxplots in 
Fig.~\ref{fig:dice_across_tissue}. The models perform best in segmenting myxoid 
and Fat tumors and perform worst in fibrous and vascular tissue types. Fat and 
myxoid tumors generally have a large homogeneous structure and are easy to 
differentiate from the background. Fibrous and vascular tissue often have a 
heterogeneous structure making it difficult to differentiate from the 
surrounding tissue. We visually confirm our observations in Fig.~\ref{fig:predicted_mask_example}. A larger list of prediction failure montages 
can be found in Appendix \ref{app:prediction_failures}.
% \reasat{Cite any literature that states it is tough also for the radiologists. 
% Is there any literature that talks about the difficulty in manually segmenting 
% the fibrous and vascular tissue? we can also report the reasons from what we 
% have seen from our perspective. We will also need to add why fat and myxoid are 
% easy to segment. Large size, homogenous structure, fat easy to differentiate in 
% T1, etc.}
%\reasat{Reporting dice scores, analyzing dice variation with class 
%distribution, anatomy distribution,  tumor size distribution, image size 
%distribution, scan resolution distribution, image intensity distribution}

\subsection{Effect of Volume}

\begin{figure*}[t!]
	\centering
	\begin{subfigure}[t]{0.33\textwidth}
		\centering
		\includegraphics[width=\columnwidth]{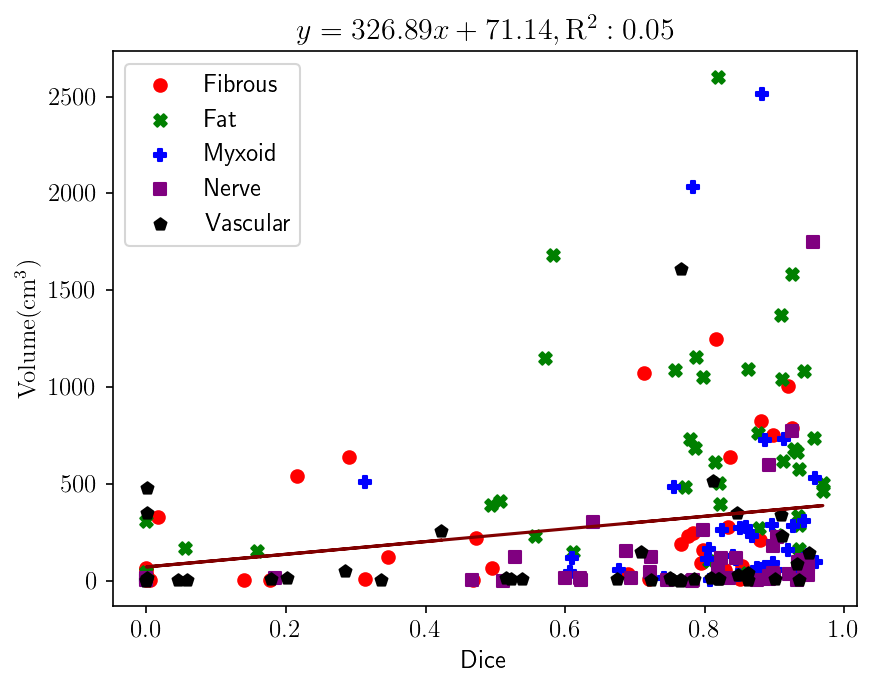}
		\caption{All tumors.}
	\end{subfigure}%
	\begin{subfigure}[t]{0.33\textwidth}
		\centering
		\includegraphics[width=\columnwidth]{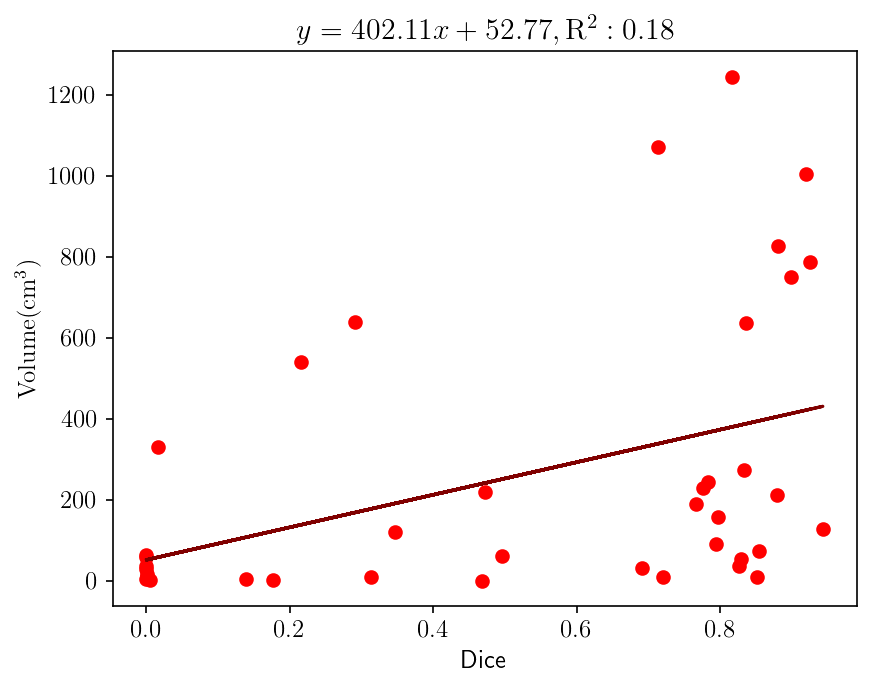}
		\caption{Fibrous.}
	\end{subfigure}
	\begin{subfigure}[t]{0.33\textwidth}
		\centering
		\includegraphics[width=\columnwidth]{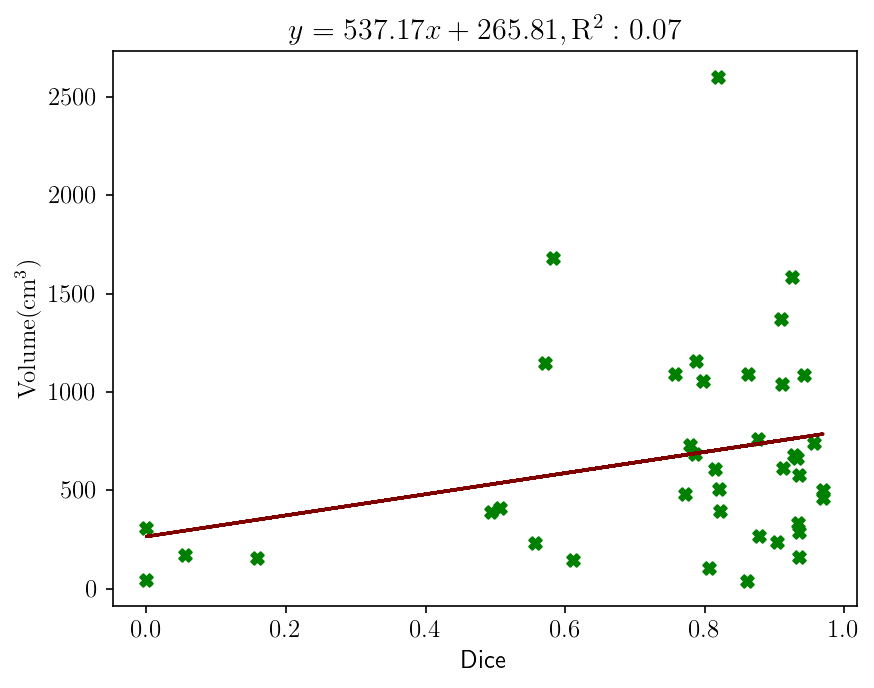}
		\caption{Fat.}
	\end{subfigure}%
    
    \centering
    \begin{subfigure}[t]{0.33\textwidth}
        \centering
        \includegraphics[width=\columnwidth]{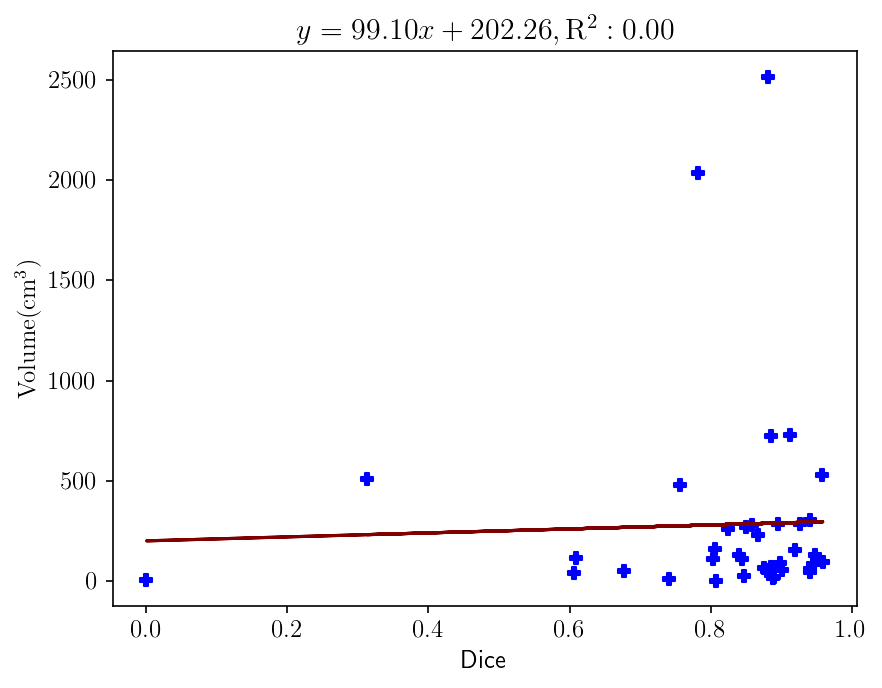}
        \caption{Myxoid.}
    \end{subfigure}%
    \begin{subfigure}[t]{0.33\textwidth}
        \centering
        \includegraphics[width=\columnwidth]{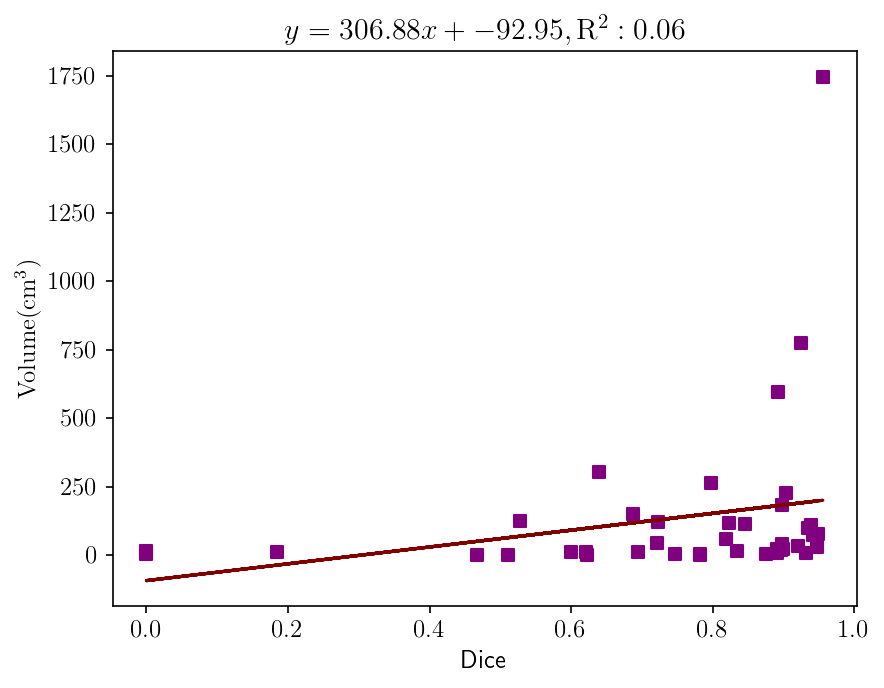}
        \caption{Nerve.}
    \end{subfigure}
    \begin{subfigure}[t]{0.33\textwidth}
        \centering
        \includegraphics[width=\columnwidth]{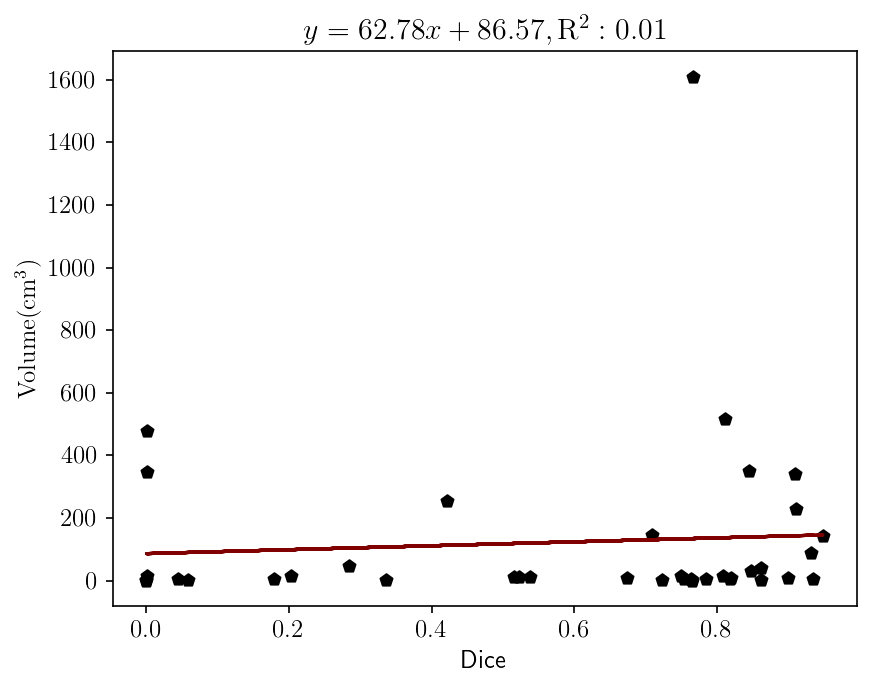}
        \caption{Vascular.}
    \end{subfigure}%
\caption{Effect of tumor volume on segmentation performance. The trend line equation and $\mathrm{R^2}$ value is shown for the tissue types. It is difficult for the model to detect small fibrous tumors.}
\label{fig:vol_vs_dice}
\end{figure*}

In general, larger tumor volumes are easier to segment for the model. Fig.~\ref{fig:vol_vs_dice} shows the performance variation of the model across 
different tissue types. Especially for the fibrous tissue type,  there's a clear 
trend of performance increase with volume. 
% \reasat{adding example image}

\subsection{Effect of Anatomy}

In Table \ref{table:dice_across_anatomy}, we report the average Dice scores 
across different tumor locations. As expected, the model does better in the 
extremities compared to the non-extremity locations due to having more sample 
representation in the training set. 
% \reasat{adding example image}

% \usepackage{multirow}

\begin{table}[htbp]
 \resizebox{\textwidth}{!}{
 \begin{tabular}{lrrrrrrr}
 \toprule
\multirow{2}{*}{Tumor Site} & \multicolumn{5}{c}{Tissue Type}                                & \multirow{2}{*}{Total Count} & \multirow{2}{*}{Avg.~Dice} \\
\cline{2-6}
                            & Fibrous & Fat   & Myxoid  & Nerve  & Vascular  &                              &                           \\
                            \midrule
thigh                       & 0.701       & 0.775    & 0.827      & 0.818     & 0.746        & 103                          & 0.774                 \\
leg                         & 0.223       & 0.893    & 0.740      & 0.612     & 0.301        & 25                           & 0.505                  \\
glute                       & -          & 0.477    & 0.815      & 0.842     & 0.356        & 19                           & 0.711                 \\
forearm                     & 0.000       & -       & -         & 0.877     & 0.723        & 13                           & 0.715                 \\
arm                         & 0.738       & 0.895    & -         & 0.540     & -           & 10                           & 0.596                  \\
hand                        & -          & -       & -         & 0.787     & 0.245        & 4                            & 0.381                 \\
shoulder                    & 0.474       & -       & -         & -        & 0.493        & 4                            & 0.485                  \\
foot                        & 0.058       & -       & -         & 0.278     & 0.718        & 3                            & 0.351                  \\
neck                        & 0.284       & -       & -         & -        & 0.663        & 3                            & 0.411                  \\
pelvis                      & 0.922       & -       & -         & 0.561     & 0.448        & 3                            & 0.595                  \\
chest wall                  & 0.173       & -       & 0.898      & -        & -           & 3                            & 0.415                  \\
abdomial wall              & 0.020       & -       & -         & -        & -           & 2                            & 0.020                  \\
axilla                      & -          & -       & -         & 0.276     & -           & 1                            & 0.276                  \\
back                        & 0.325       & -       & -         & -        & -           & 1                            & 0.325                  \\
flank                       & -          & -       & -         & -        & 0.148        & 1                            & 0.148                  \\
\bottomrule
\end{tabular}
}
\caption{Average Dice score distribution for different anatomical locations. Missing values mean that the dataset did not contain a tumor of that type in that particular anatomical location.}
\label{table:dice_across_anatomy}
\end{table}

% \begin{table}[htbp]
% \centering
% \begin{tabular}{cll}
% \toprule
% \textbf{Tumor Site} & \textbf{Count} & \textbf{Avg Dice} \\
% \midrule
% thigh            & 103            & 0.774126   \\
% leg              & 25             & 0.504854   \\
% glute            & 19             & 0.711305   \\
% forearm          & 13             & 0.715211   \\
% arm              & 10             & 0.595697   \\
% shoulder         & 7              & 0.485172   \\
% hand             & 4              & 0.380507   \\
% pelvis           & 4              & 0.595086   \\
% chest wall       & 3              & 0.414867   \\
% foot             & 3              & 0.350981   \\
% neck             & 3              & 0.410597   \\
% abdominal wall   & 2              & 0.020027   \\
% axilla           & 1              & 0.276095   \\
% back             & 1              & 0.325348   \\
% flank            & 1              & 0.148313   \\
% \bottomrule

% \end{tabular}
% \end{table}

\subsection{Effect of Tumor Intensity}

\begin{figure*}[htbp]
	\centering
	\begin{subfigure}[t]{0.33\textwidth}
		\centering
		\includegraphics[width=\columnwidth]{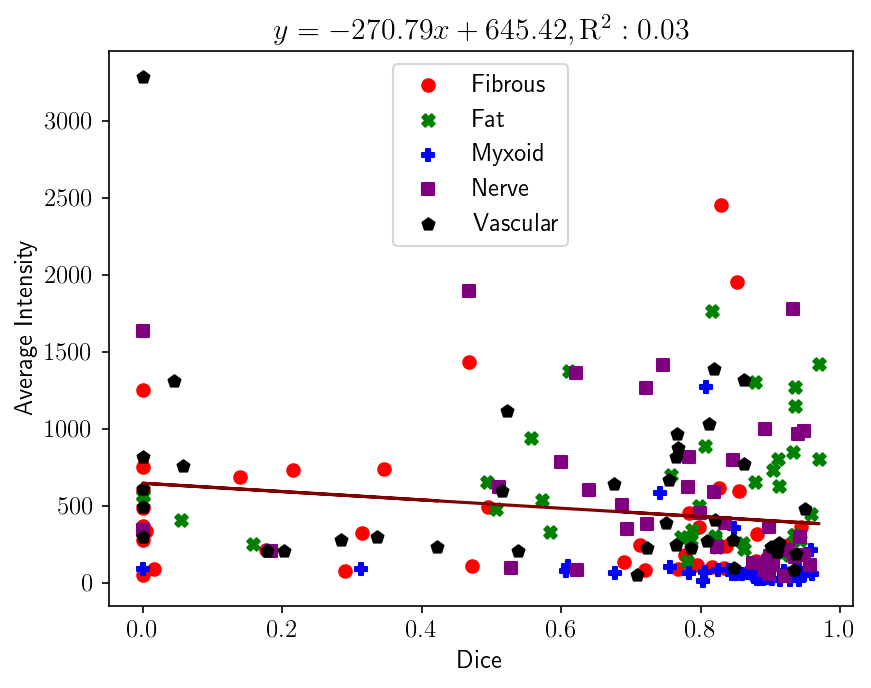}
		\caption{All tumors.}
	\end{subfigure}%
	\begin{subfigure}[t]{0.33\textwidth}
		\centering
		\includegraphics[width=\columnwidth]{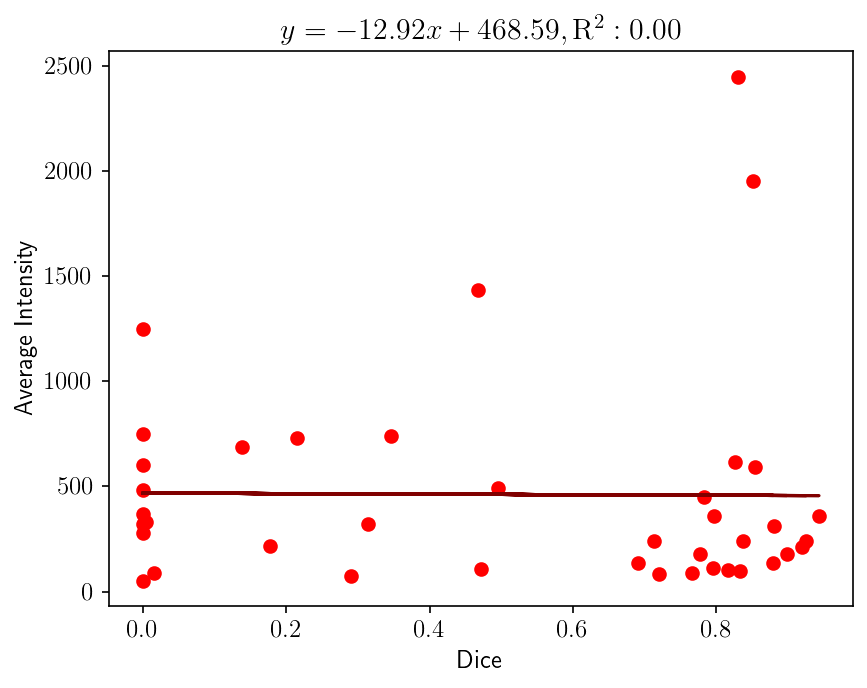}
		\caption{Fibrous.}
	\end{subfigure}
	\begin{subfigure}[t]{0.33\textwidth}
		\centering
		\includegraphics[width=\columnwidth]{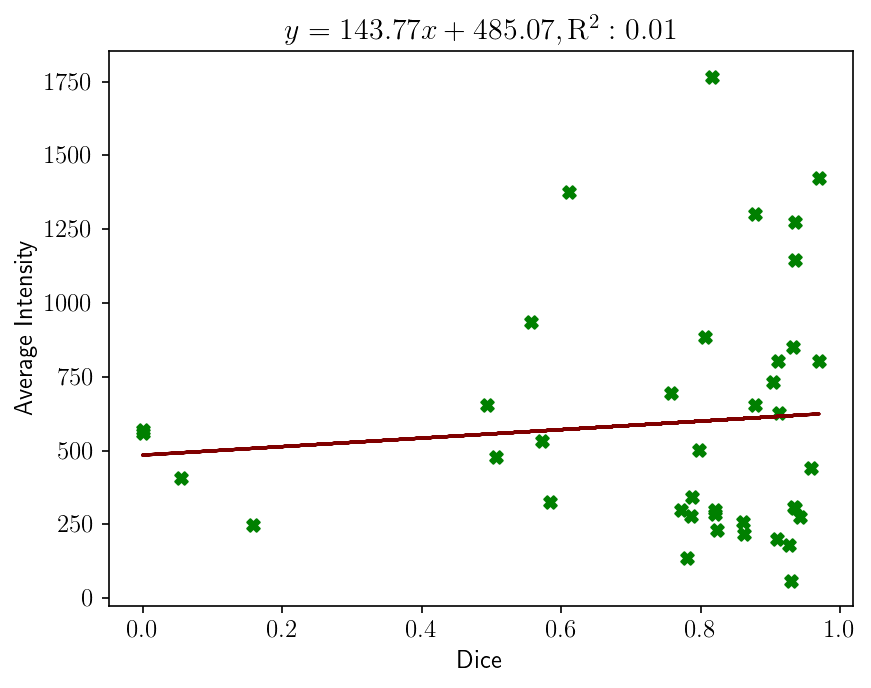}
		\caption{Fat.}
	\end{subfigure}%
    
    \centering
    \begin{subfigure}[t]{0.33\textwidth}
        \centering
        \includegraphics[width=\columnwidth]{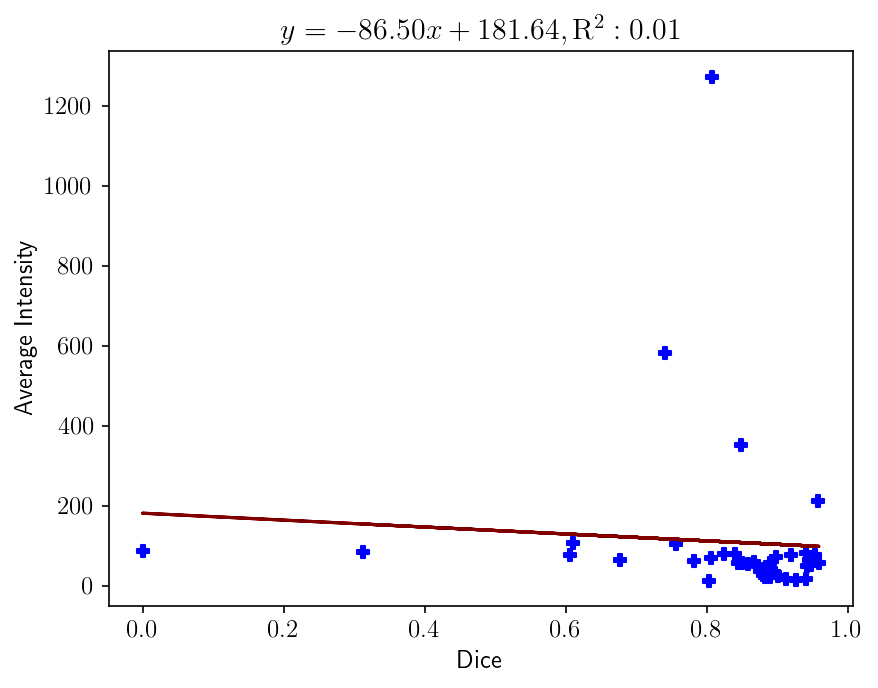}
        \caption{Myxoid.}
    \end{subfigure}%
    \begin{subfigure}[t]{0.33\textwidth}
        \centering
        \includegraphics[width=\columnwidth]{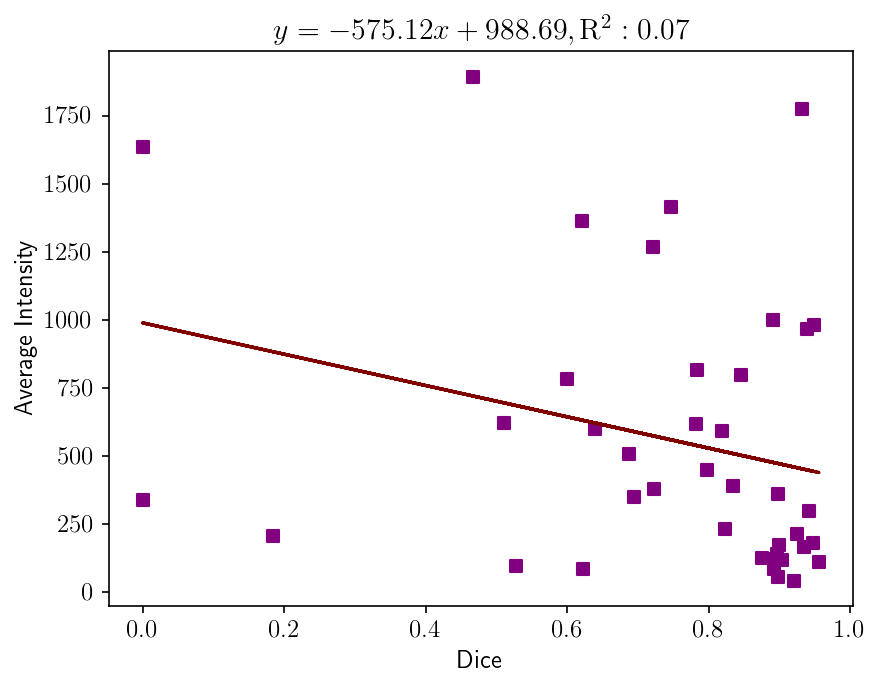}
        \caption{Nerve.}
    \end{subfigure}
    \begin{subfigure}[t]{0.33\textwidth}
        \centering
        \includegraphics[width=\columnwidth]{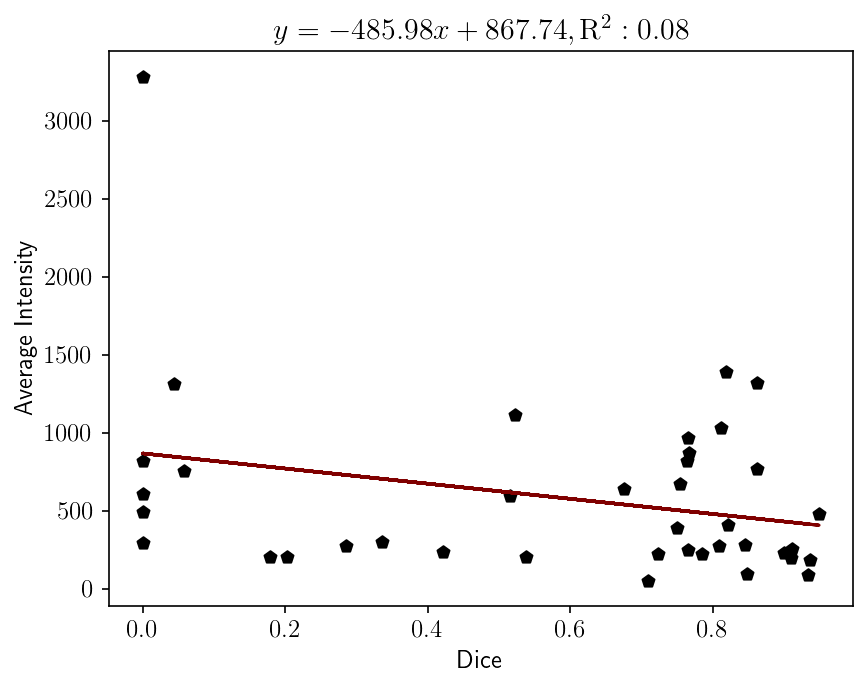}
        \caption{Vascular.}
    \end{subfigure}%

\caption{Effect of average tumor intensity (T1 image) on segmentation 
	performance. For the fat tumors, there is a weak correlation of dice 
	improvement with the increase of intensity.
	%Scatter plot of tumor volume vs dice score shows there is a general 
	%tendency of dice improvement with larger tumors.
	}

\label{fig:avg_int_t1_vs_dice}
\end{figure*}

There is a weak positive correlation between average tumor intensity in the T1 
image and dice scores for Fat tissue (Fig.~\ref{fig:avg_int_t1_vs_dice}). This 
is expected as lipid content is more distinguishable in the T1 image. However, 
for the rest of the tissue types, an increase in brightness has a no positive effect 
on tumor localization as those tissues are generally darker in the T1 image. 
% \reasat{what is the tissue content that makes detection tougher?}

\subsection{Suggestion on Future Data Collection}

Although our dataset is large enough to train a state-of-the-art model, our result analysis 
suggests a dire need for a larger dataset to capture more diversity of 
tissue structures. Although we have set up a balanced dataset in terms of tissue 
types, there is an imbalance in terms of anatomy. In Table 
\ref{table:anatomy_tissue_distribution}, we see fibrous and vascular tissue 
types have diverse anatomical representations. As observed in Table 
\ref{table:dice_across_tissue} and Table \ref{table:dice_across_anatomy}, these 
tissue types have poorer Dice scores compared to myxoid and Fat, even in 
the most common anatomical locations such as the extremities. The lesser 
representation of these difficult tissue types adds up with the challenging 
visual characteristics (small size, unclear tumor boundaries) and worsens the 
learning capability of the model. Future iterations of this dataset should focus 
on collecting more of these less representative and diverse tissue types.

% \reasat{Implement special training procedures to address small volume segmentation, explore weighted loss, focal loss, etc. Also, adopt self-supervised learning mechanisms to learn a rich encoder representation and explore transfer learning, finetuning, and parameter-efficient fine-tuning approaches.}

\section{Conclusion}

\label{sec_conclusion}
In this work, we have described the creation of an MSTT dataset. We 
have trained a segmentation model on this dataset and benchmarked its 
performance on a publicly available dataset which achieves state-of-the-art result on MSTT segmentation. Results show that the segmentation models 
work well for the Fat, myxoid, and nerve tumors but struggle to segment tumors 
on fibrous and Vvscular tumors. The segmentation model is sensitive to the 
volume of the tissue as well as the tumor location. Although this is the largest 
tumor segmentation dataset created, the size of the dataset needs to be 
increased further to make the segmentation models more robust. Special priority needs to 
be given to the tumors with fibrous and vascular tissue types as they have 
diverse anatomical locations and have challenging visual characteristics 
compared to Myxoid, Fat, and Nerve tissue.
% \bibliographystyle{} 
\bibliography{cas-refs}

\appendix

\section{Training Parameters}
\label{app:training_params}
List of augmentations used from the \texttt{albumentation} package:
\begin{itemize}
\itemsep=-0.3em
	\item  \texttt{PadIfNeeded(p=0.5)},
	\item  \texttt{RandomCrop(p=0.5)},
	\item  \texttt{RandomRotate90(p=0.5)},
	\item  \texttt{VerticalFlip(p=0.5)}
	\item  \texttt{HorizontalFlip(p=0.5)},
	\item  \texttt{RandomGamma(gammaLimit=(50, 200), p=0.5)},
	\item  \texttt{RandomBrightnessContrast(p=0.5)},
	\item  \texttt{GaussianBlur(p=0.5)},
	\item  \texttt{MotionBlur(p=0.5)},
	\item  \texttt{GridDistortion(numSteps=5, distortLimit=0.3, p=0.5)},
	\item  \texttt{Normalize(mean=0, std=1)}.
\end{itemize}

List of model parameters
\begin{itemize}
\itemsep=-0.3em
	\item Learning rate: 1e-4
	\item Batch size: 16
	\item Epochs: 5
\end{itemize}

\section{Prediction Failures}
\label{app:prediction_failures}

A list of images where the segmentation model has failed.

\begin{figure}
    \centering
    \includegraphics[width=\textwidth]{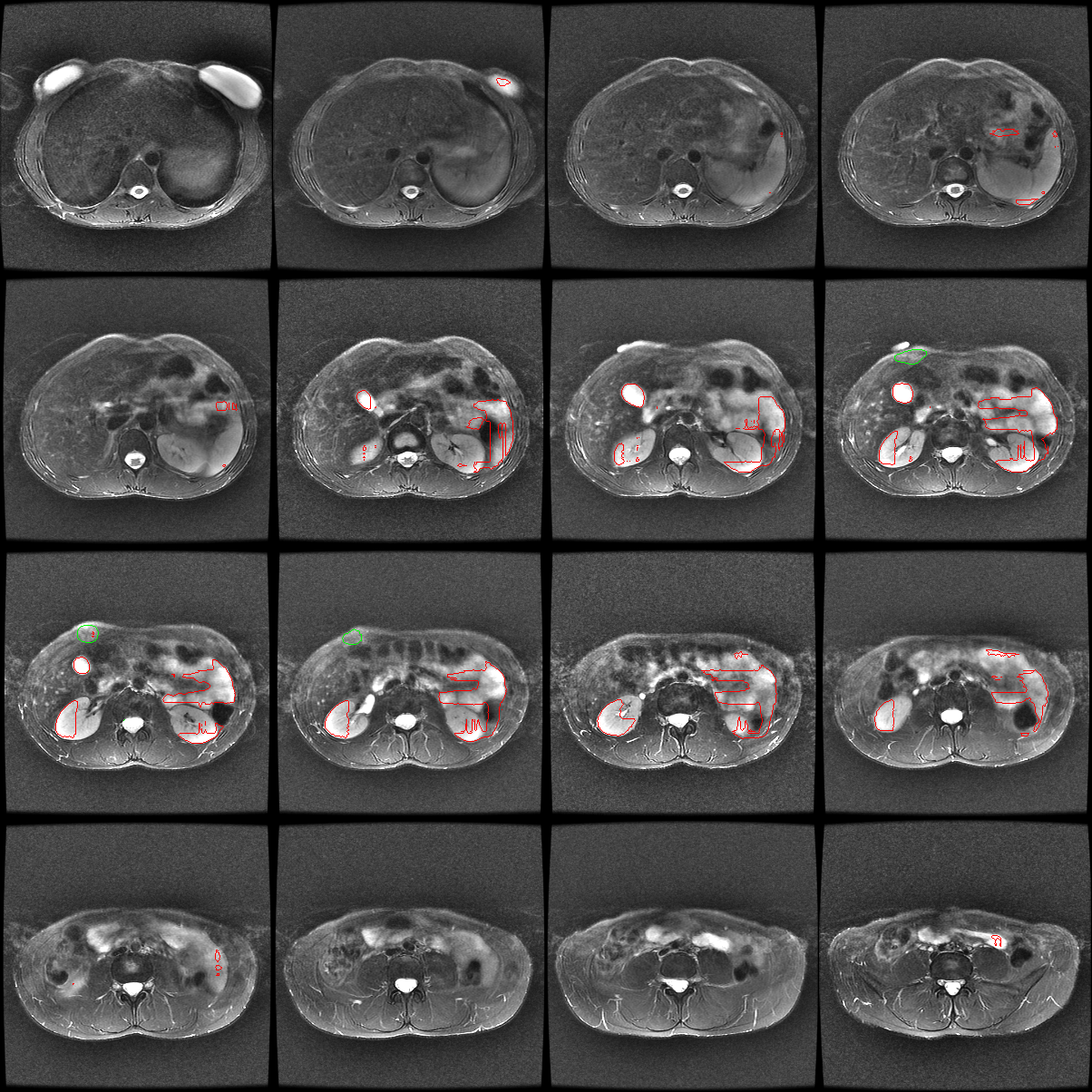}
    \caption{Fibrous tumor in the abdominal wall. There are confounding structures that create false positives. The tumor tissue gets mixed up with surrounding tissue creating weak boundaries. (Green contour is ground truth, red contour is prediction)}
    \label{fig:prediction_failures_1}
\end{figure}

\begin{figure}
    \centering
    \includegraphics[width=\textwidth]{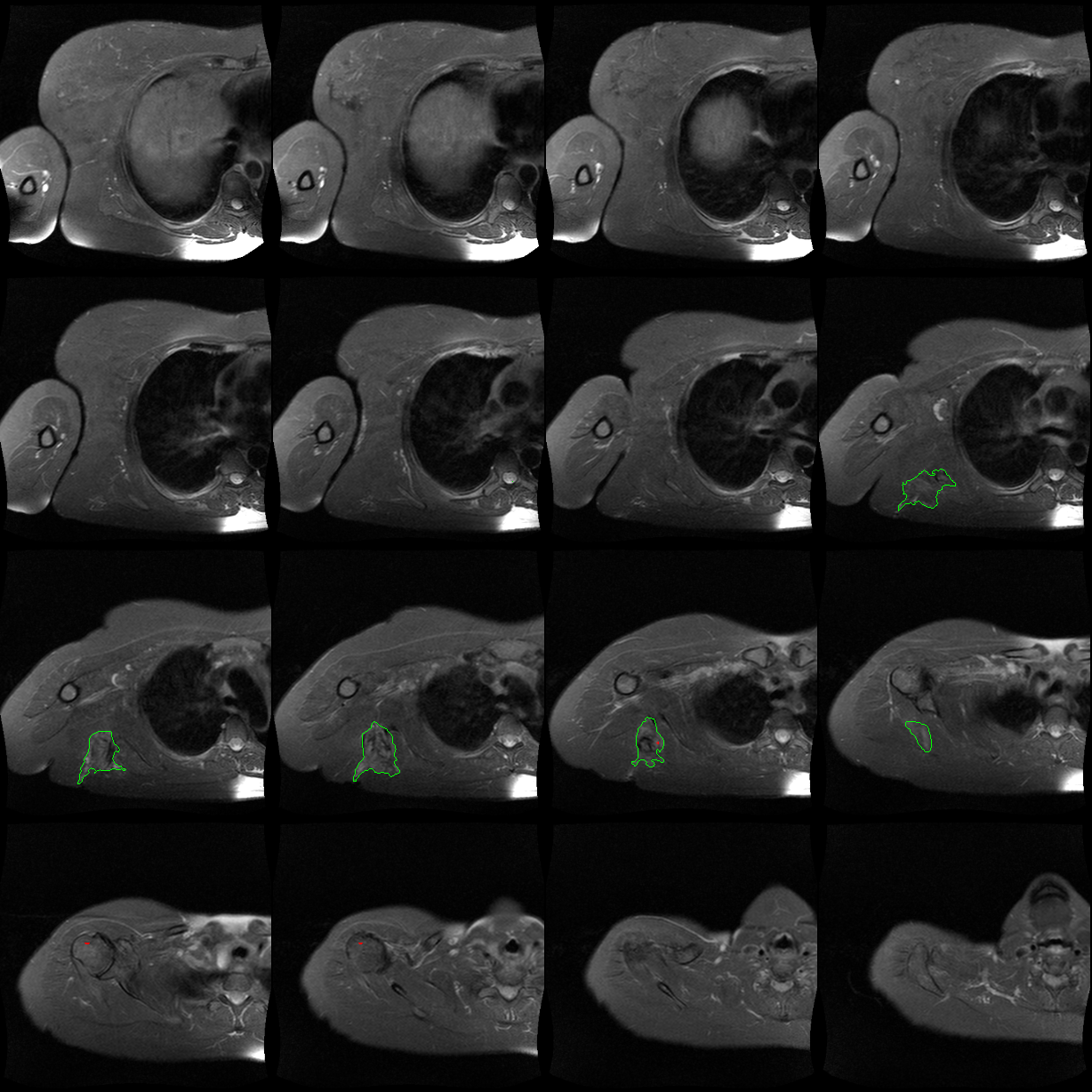}
    \caption{Fibrous tumor in the chest wall. The tumor tissue is not separable from surrounding tissue in terms of intensity and therefore has weak boundaries. (Green contour is ground truth, red contour is prediction)}
    
    \label{fig:prediction_failures_2}
\end{figure}

\begin{figure}
    \centering
    \includegraphics[width=\textwidth]{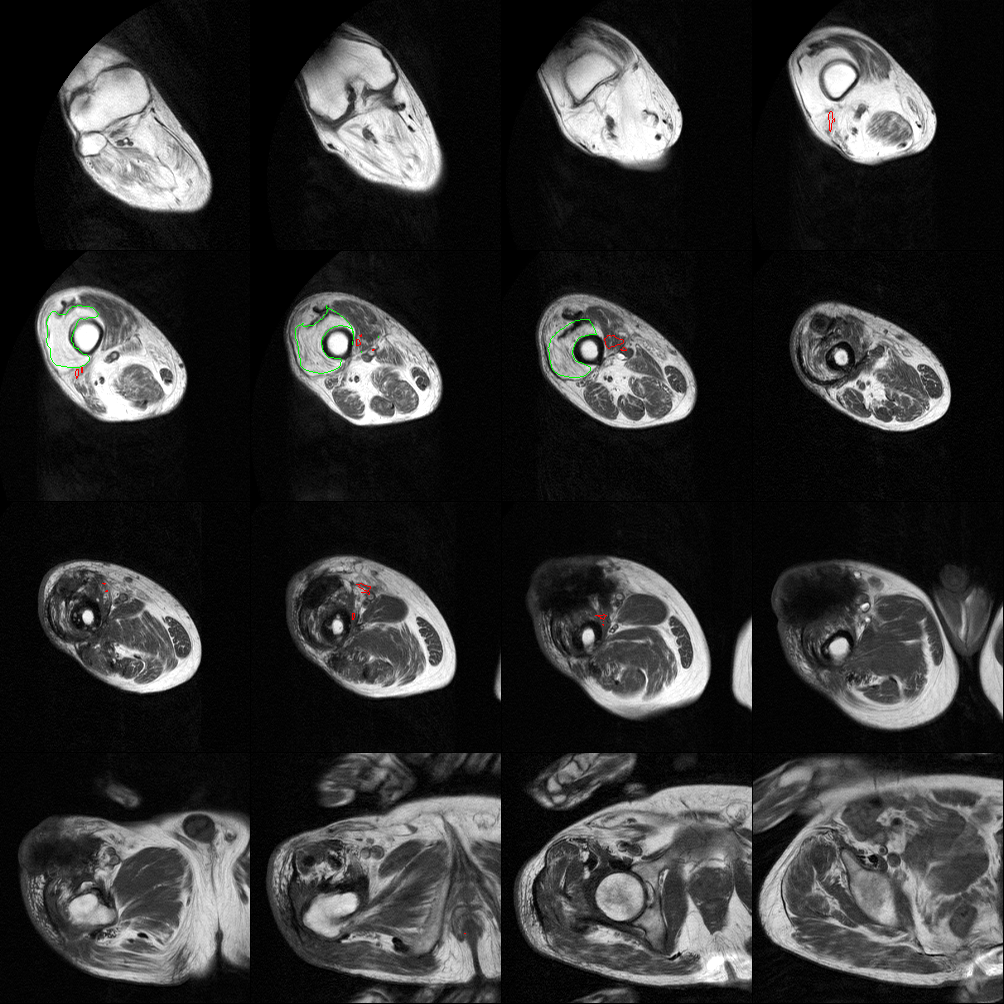}
    \caption{Fat tumor in the thigh. The tumor tissue is not separable from surrounding tissue in terms of intensity and therefore has weak boundaries. (Green contour is ground truth, red contour is prediction)}
    \label{fig:prediction_failures_3}
\end{figure}
\begin{figure}
    \centering
    \includegraphics[width=\textwidth]{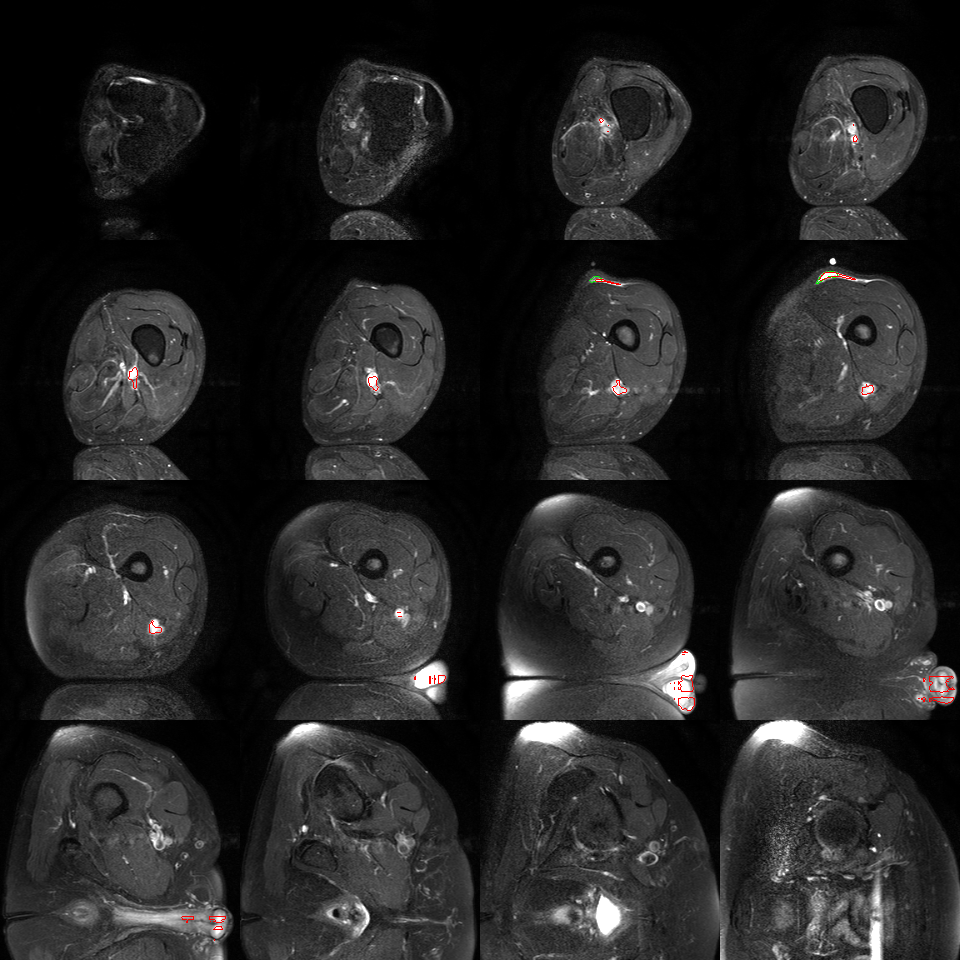}
    \caption{Myxoid tumor in the thigh. The testicles behave as confounding tissue creating false positives. (Green contour is ground truth, red contour is prediction)}
    \label{fig:prediction_failures_4}
\end{figure}
\begin{figure}
    \centering
    \includegraphics[width=\textwidth]{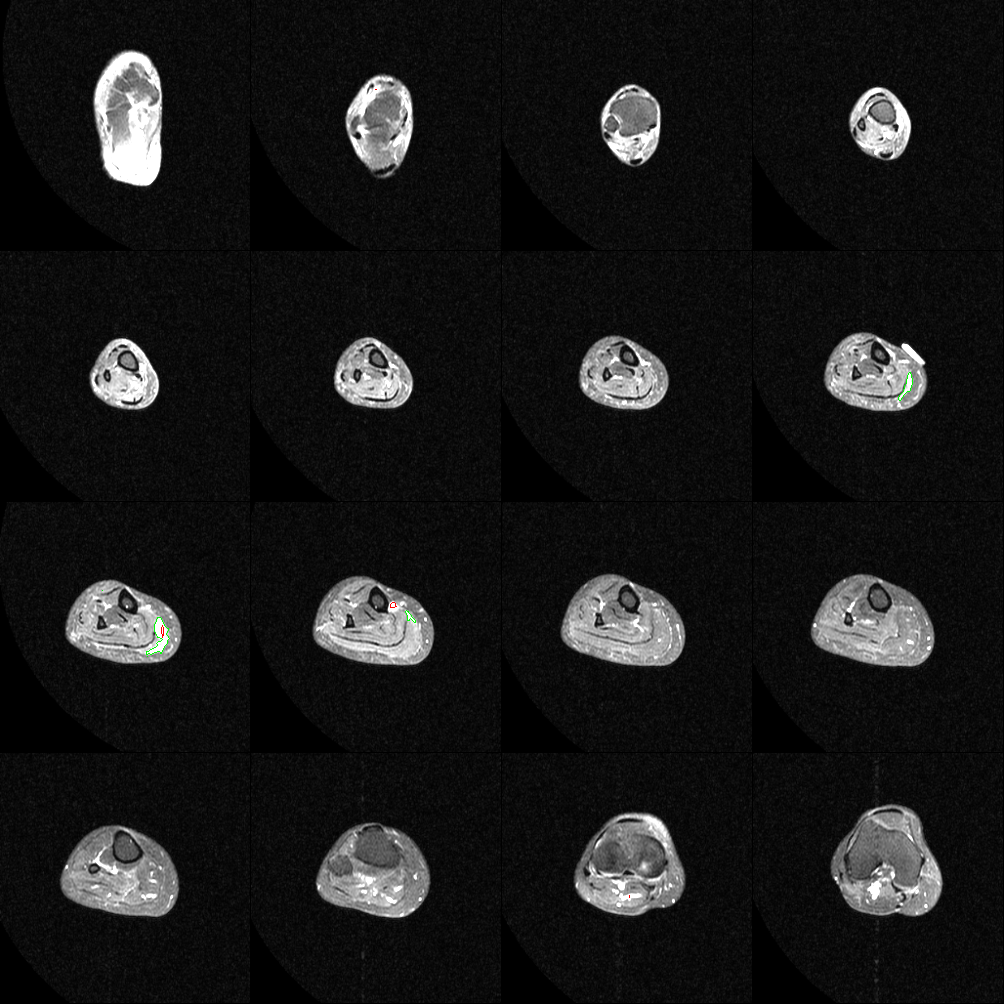}
    \caption{Nerve tumor in the leg. The tumor tissue is separable, but the image has artifacts confounding the model. (Green contour is ground truth, red contour is prediction)}
    \label{fig:prediction_failures_5}
\end{figure}
\begin{figure}
    \centering
    \includegraphics[width=\textwidth]{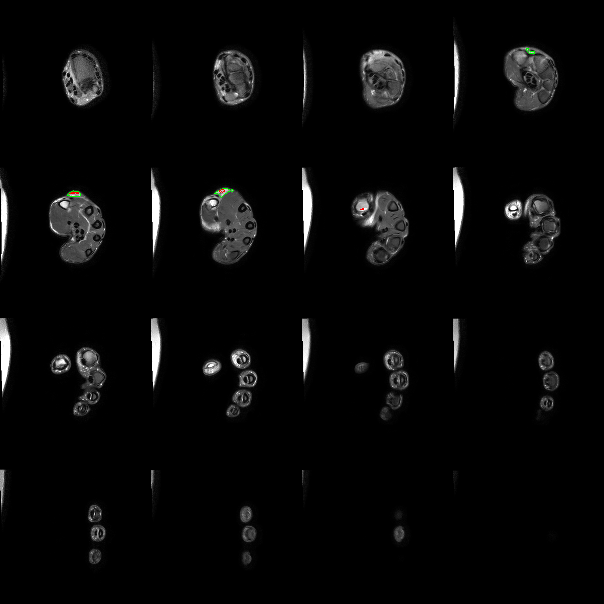}
    \caption{Vascular tumor in the hand. This is a rare anatomical structure with very few representations in the dataset. (Green contour is ground truth, red contour is prediction)}
    
    \label{fig:enter-label}
\end{figure}
\begin{figure}
    \centering
    \includegraphics[width=\textwidth]{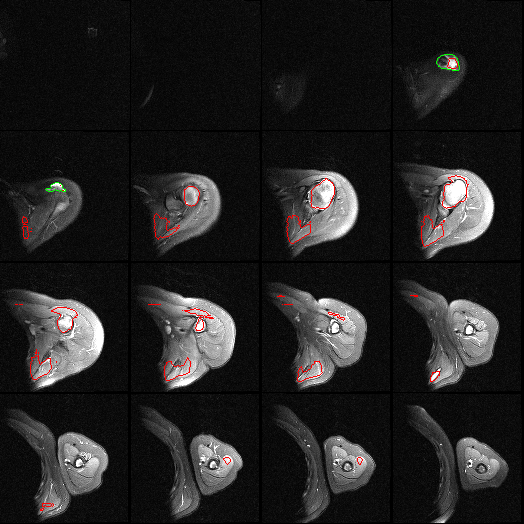}
    \caption{Vascular tumor in the shoulder. The tumor tissue has a heterogeneous structure, and the area with less intensity is not captured. Also, the bone confounds the model. (Green contour is ground truth, red contour is prediction)}
    
    \label{fig:prediction_failures_6}
\end{figure}

\begin{figure}
    \centering
    \includegraphics[width=\textwidth]{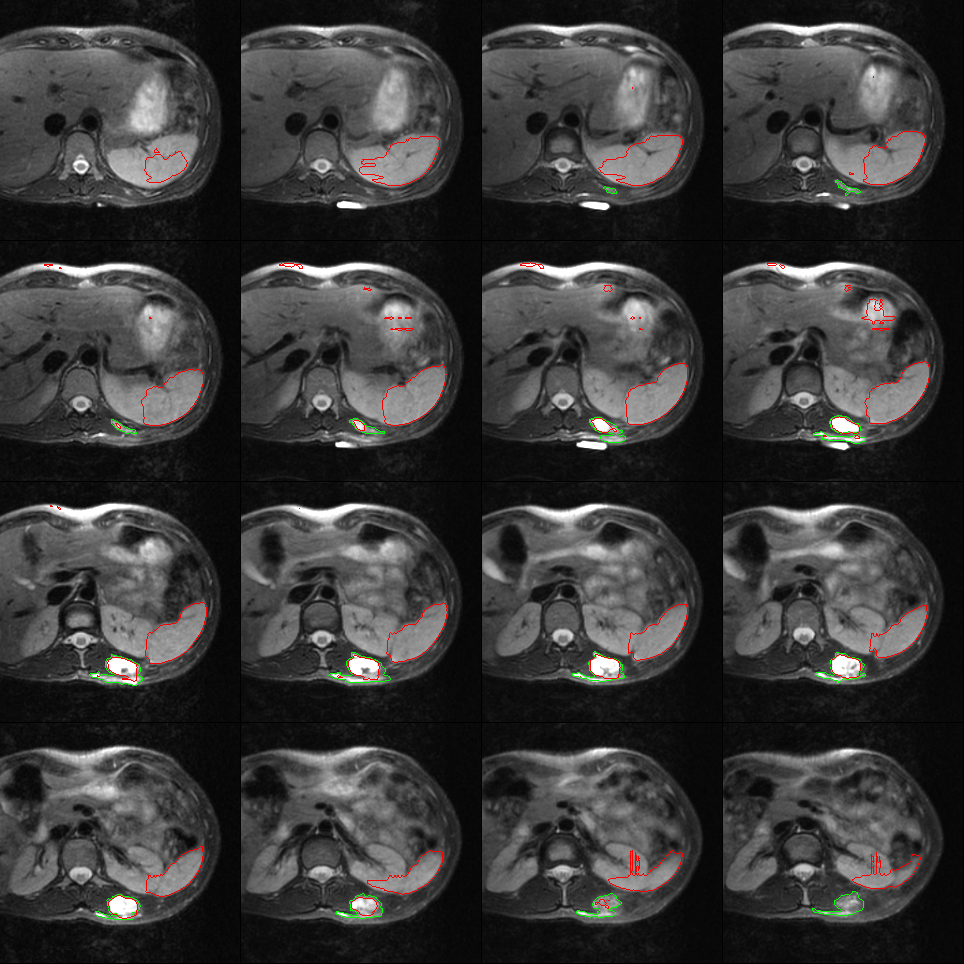}
    \caption{Vascular tumor in the flank. This is a rare anatomy, and the confounding structures create false positives. (Green contour is ground truth, red contour is prediction)}    \label{fig:prediction_failures_7}
\end{figure}

\end{document}